\documentclass[aps,floats]{revtex4}
\usepackage{amsmath,amssymb}
\usepackage{graphicx,epsfig}

\begin{document}
\bibliographystyle {plain}

\def\oppropto{\mathop{\propto}} 
\def\opsimeq{\mathop{\simeq}}
\def\opoverderline{\mathop{\overline}}
\def\operarrow{\mathop{\longrightarrow}}
\def\opsim{\mathop{\sim}}

\def\fig#1#2{\includegraphics[height=#1]{#2}}
\def\figx#1#2{\includegraphics[width=#1]{#2}}


\title{Critical behavior of interfaces in disordered Potts
ferromagnets : \\
statistics of free-energy, energy and interfacial adsorption  } 


 \author{ C\'ecile Monthus and Thomas Garel }
  \affiliation{Service de Physique Th\'{e}orique, CEA/DSM/SPhT\\
 Unit\'e de recherche associ\'ee au CNRS\\
 91191 Gif-sur-Yvette cedex, France}

\begin{abstract}

A convenient way to study phase transitions of finite spins systems
of linear size $L$ is to fix boundary conditions that impose the presence 
of a system-size interface. In this paper, we study the statistical properties
of such an interface in a disordered Potts ferromagnet in dimension
$d=2$
within Migdal-Kadanoff real space renormalization.
We first focus on the interface free-energy and energy 
to measure the singularities of the average and random contributions,
as well as the corresponding histograms,
 both in the low-temperature phase and at criticality.
We then consider the critical behavior
of the interfacial adsorption of non-boundary states.
Our main conclusion is that all singularities involve the 
correlation length $\xi_{av}(T) \sim (T_c-T)^{-\nu}$ appearing in the average 
free-energy $\overline{F} \sim (L/\xi_{av}(T))^{d_s}$
of the interface of dimension $d_s=d-1$, except for
the free-energy width $\Delta F \sim (L/\xi_{var}(T))^{\theta}$
that involves the droplet exponent $\theta$ 
and another correlation length $\xi_{var}(T)$ which diverges
more rapidly than $\xi_{av}(T)$.
We compare
with the spin-glass transition in $d=3$, 
where $\xi_{var}(T)$ is the
'true' correlation length, and where the interface energy presents 
unconventional scaling with a chaos critical exponent $\zeta_c>1/\nu$
[Nifle and Hilhorst, Phys. Rev. Lett. 68, 2992  (1992)].
The common feature is that in both cases,
the characteristic length scale $L_{ch}(T)$ associated with the chaotic
nature of the low-temperature phase,
diverges more slowly than the correlation length.
\end{abstract}

\maketitle

\section{ Introduction }

 \subsection{ Critical properties of interfaces in pure systems }

The critical points of statistical physics models
are usually discussed in terms of bulk properties.
However, it is also interesting to study how a critical system
reacts when boundary conditions impose the presence 
of a system-size interface. For instance in an Ising ferromagnet
defined on a cube of volume $L^d$, on may impose
the $(+)$ phase on the left boundary and the $(-)$ 
phase on the right boundary,
with periodic boundary conditions in the other directions.
The study of interfaces between coexisting phases near criticality
 has a long history \cite{Widom}.
The most important property is that the free-energy $F^{inter}$
associated to the interface 
is proportional below $T_c$ to its area $L^{d_s}$ with $d_s=d-1$
\begin{eqnarray}
{\rm Pure \ Ferromagnets : } \ \ F^{inter}(L,T<T_c) \simeq f_0(T) L^{d_s} +...
\label{finterpure}
\end{eqnarray}
where the so-called interfacial tension $f_0(T)$ vanishes 
at criticality as a power-law \cite{Widom}
\begin{eqnarray}
 f_0(T) \opsimeq_{T \to T_c^-} (T_c-T)^{\mu}
\label{sigmamu}
\end{eqnarray}
Finite-size scaling implies that near criticality, the interface free-energy 
$F^{inter}$ should only depend on the ratio $L/\xi(T)$
between the system linear size $L$ and the correlation length
$\xi(T) \propto (T_c-T)^{-\nu}$
\begin{eqnarray}
F^{inter}(L,T<T_c) \propto \left( \frac{L}{\xi(T)} \right)^{d_s} +...
\label{interfacialfss}
\end{eqnarray}
This is similar to the requirement that the singular part 
of the bulk free-energy should scale as $F^{bulk} \sim (L/\xi(T))^d
\sim (T_c-T)^{\nu d} L^d $. The identification
with the definition $F^{bulk} \sim (T_c-T)^{2-\alpha} L^d$ 
in terms of the specific heat exponent $\alpha$ yields
the hyperscaling relation $2-\alpha= \nu d$. So the exponent $\mu$ of
the interfacial tension $f_0(T)$ satisfies the Widom relation 
\cite{Widom}
\begin{eqnarray}
 \mu= \nu d_s = \nu (d-1) 
\label{murelation}
\end{eqnarray}
Exactly at criticality, the free-energy becomes of order 
$F^{inter}(L,T_c) \sim O(1)$,
whereas the energy and entropy grow as $E^{inter}(L,T_c) \sim L^{1/\nu}$.
Beyond these thermodynamic properties,
the interest into interfaces was recently revived 
by the discovery \cite{schramm}
that some two dimensional critical interfaces
are fractal curves which can be constructed via
 Stochastic Loewner Evolutions (SLEs) reviewed in \cite{sle}.
Accordingly, the fractal dimensions of spin cluster boundaries 
of various two-dimensional spin models 
have been recently measured via Monte Carlo simulations in
\cite{gamsa,santa}.

Whenever the system under study presents more than two phases,
such as the Potts model considered in the this paper,
a system-size interface between states 1 and 2 tends to
produce a net adsorption of any non-boundary state, called
 state 3 here. This phenomenon of interfacial adsorption has been much studied
in various pure models \cite{selke} with the following conclusions.
The excess of state '3' due to the presence of a (1:2) interface
with respect to the case (1:1) with no interface, defined as
\begin{eqnarray}
 N_{nb} \equiv \sum_{i} 
\left( < \delta_{\sigma_i,3}>_{1:2} - < \delta_{\sigma_i,3}>_{1:1}\right)
\label{defnb}
\end{eqnarray}
is proportional to the area $L^{d_s}$ of the interface for $T<T_c$
\begin{eqnarray}
 N_{nb}(L,T<T_c) \sim w_0(T) L^{d_s}
\label{defnbbelow}
\end{eqnarray}
Finite-size scaling argument yields that the coefficient $w_0(T)$
diverges at criticality as \cite{selke}
\begin{eqnarray}
 w_0(T) \oppropto_{T \to T_c^-} (T_c-T)^{\beta- \nu}
\label{singw0}
\end{eqnarray}
where $\nu$ is the correlation length introduced above,
and where $\beta$ is the order parameter exponent.
 This means that at criticality, the adsorption of non-boundary states
$N_{nb}$ scales as the global order parameter $M = (T_c-T)^{\beta} L^d$.
\begin{eqnarray}
 N_{nb}(L,T_c) \sim M(L,T_c) \sim L^{d- \frac{\beta}{\nu}}
\label{nbtc}
\end{eqnarray}

 \subsection{ Properties of interfaces 
below $T_c$ in disordered systems }

In the field of disordered systems such as spin-glasses where 
the order parameter
of the low-temperature phase is more 
complicated than in pure systems,
it turns out that the properties of interfaces are very convenient
to characterize the low-temperature phase via a so-called
droplet exponent $\theta$ \cite{Fis_Hus,heidelberg}
\begin{eqnarray}
{\rm Spin-glass : } \ \  F^{inter}(L,T<T_c)  = \Upsilon(T) L^{\theta} u_F +...
\label{thetasg}
\end{eqnarray}
where $ \Upsilon(T)$ is a generalized 'stiffness' modulus
 and where $u_F$ is a random variable of order $O(1)$.
The exponent $\theta$ is expected to satisfy
 the bound $\theta \leq (d-1)/2$ \cite{Fis_Hus}
The interface is expected to have a non-trivial fractal dimension
$d_s$ with $d-1 \leq d_s \leq d$ in the whole
low-temperature phase \cite{Fis_Hus}. 
This fractal dimension $d_s$  governs the energy and the entropy of
the interface \cite{Fis_Hus} 
\begin{eqnarray}
{\rm Spin-glass : } \ \ 
  E^{inter}(L,T<T_c) && = \sigma(T) L^{\frac{d_s}{2}} u_E+...  \\
\nonumber
T S^{inter}(L,T<T_c) && = \sigma(T) L^{\frac{d_s}{2}} u_E+
\label{enersg}
\end{eqnarray}
One actually expects the strict inequality 
$\theta < \frac{d_s}{2}$, so that the optimized free-energy
of Eq \ref{thetasg} is a near cancellation 
of much larger energy and entropy contributions of Eq.
 \ref{enersg}. This is at the origin of the
sensitivity of disordered systems to temperature changes
or disorder changes, called 'chaos' in this context 
\cite{Fis_Hus,heidelberg,Ban_Bray,muriel,thill} :
roughly speaking, the chaos exponent $\zeta=\frac{d_s}{2}- \theta$ governs
the length scale $L^* \sim \epsilon^{-1/\zeta}$ above which
a small perturbation $\epsilon$ in the temperature
or in the disorder will change the state of the system.

For non-frustrated disordered systems 
such as ferromagnetic spin models in dimension $d$,
the interface below $T_c$ is expected to be described
by a directed manifold of dimension $d_s=(d-1)$ in a random medium.
In particular, in two-dimensional
disordered ferromagnets,  
the one-dimensional interface 
is described by the directed polymer model \cite{Hus_Hen}.
For this model, a droplet scaling theory
 has been developed \cite{Fis_Hus_DP}
in direct correspondence with the droplet theory of 
spin-glasses \cite{Fis_Hus} summarized above.
In particular, the free-energy of the interface reads
\begin{eqnarray}
{\rm Random \ Ferromagnets  : } \ \ \ \ 
F^{inter}(L,T<T_c)  = f_0(T) L^{d_s} +
 \Upsilon(T) L^{\theta(d)} u_F +...
\label{thetadp}
\end{eqnarray}
 with a droplet exponent $\theta$
which is exactly known to be $\theta(d=2)=1/3$
on the two-dimensional lattice
\cite{Hus_Hen_Fis,Kar,Joh}.
The energy and the entropy of the interface
reads \cite{Fis_Hus_DP,entropy}
\begin{eqnarray}
\label{enerdp}
{\rm Random \ Ferromagnets  : } \ \ \ \ 
E^{inter}(L,T<T_c) && = e_0(T) L^{d_s} + \sigma(T) L^{\frac{d_s}{2}} u_E +...
\\
\nonumber
T S^{inter}(L,T<T_c) &&  = T s_0(T) L^{d_s} + \sigma(T) L^{\frac{d_s}{2}} u_E +...
\end{eqnarray}
where the fluctuating term has again a bigger exponent $\frac{d_s}{2} > \theta$
that the fluctuating term of the free-energy of Eq. \ref{thetadp}.
As a consequence, the interface is again very sensitive
to temperature or disorder changes with the chaos exponent
 $\zeta=d_s/2-\theta$. In particular in dimension $d=2$, where the interface
is a directed polymer in dimension $1+1$, the chaos exponent is
exactly known $\zeta=1/2-1/3=1/6$
\cite{Fis_Hus_DP,zhang,feigelman,shapir}.

 \subsection{
 Properties of interfaces at criticality in disordered systems }

At criticality,
the interface free-energy is expected to be a random variable $u_{F_c}$
of order $O(1)$ 
\begin{eqnarray}
  F^{inter}(L,T_c) =   u_{F_c}+... 
\label{freecriti}
\end{eqnarray}
For the spin-glass case, the interface free-energy 
of Eq. \ref{thetasg}
is expected to scale as $(L/\xi(T))^{\theta}$
in terms of the diverging correlation length $\xi(T) \sim (T_c-T)^{-\nu}$,
so that the critical exponent governing
the vanishing of $\Upsilon(T)$ is \cite{Fis_Hus}
\begin{eqnarray}
 \Upsilon(T) \sim (T_c-T)^{\nu \theta}
\label{upsiloncriti}
\end{eqnarray}
which is the analog of Widom scaling relation for ferromagnets
(Eq. \ref{murelation}).

 For the energy of the interface, two possibilities have
been described in the literature :

(i) in the first scenario described in \cite{Fis_Hus},
the critical behavior follows the usual finite-size scaling
forms in terms of the diverging correlation
 length $\xi(T) \sim (T_c-T)^{-\nu}$. More precisely, the singular part
of the energy or entropy is assumed to be of order $1/(T_c-T)$
on the scale $\xi(T)$, so that the coefficient $\sigma(T)$ in Eq. \ref{enersg}
presents the following singularity
\begin{eqnarray}
{\rm SG \ with \ 'Conventional' \ critical \ point : } \ \ 
  \sigma(T) \opsimeq_{T \to T_c^-} \frac{1}{T_c-T} \left( \frac{1}{\xi(T)}
 \right)^{\frac{d_s}{2}} 
\label{enersgcritiusualfss}
\end{eqnarray}
Equivalently, one then obtains  
 the following 'conventional random critical' behavior 
exactly at criticality \cite{Fis_Hus}
\begin{eqnarray}
{\rm SG \ with \ 'Conventional' \  critical \ point : } \ \ 
  E^{inter}(L,T_c) =  L^{\frac{1}{\nu}} u_{E_c}+... 
\label{enersgcriticonventional}
\end{eqnarray}
where $u_{E_c}$ is a random variable of order $O(1)$.
In our recent study of the directed polymer delocalization transitions
on hierarchical lattices with $b=5$ \cite{diamondcritipolymers},
we have found that the energy and entropy are governed
by the 'conventional' critical behaviors of Eqs \ref{enersgcritiusualfss} and
\ref{enersgcriticonventional}.

(ii) 
however in \cite{muriel,thill}, it has been found that a new
exponent $\zeta_c$ called the 'critical chaos exponent'
can govern the response to disorder perturbations of spin-glasses 
at criticality, 
provided the inequality $\zeta_c>1/\nu$
is satisfied. We refer to \cite{muriel,thill}
for a detailed description of these chaos properties.
Here, we will only mention an important consequence  
for the interface : it has been argued in \cite{muriel,thill} that
this new exponent $\zeta_c$ should govern the scaling of the interface
energy at criticality  
\begin{eqnarray}
{\rm SG \ with 'Chaos \ critical \ exponent' : } \ \ 
  E^{inter}(L,T_c) =  L^{\zeta_c} u_{E_c}+... 
\label{enersgcritichaos}
\end{eqnarray}
in contrast with Eq. \ref{enersgcriticonventional}.
As a final remark on spin-glasses, let us mention
that in $d=2$ where there is no spin-glass phase ($T_c=0$),
recent studies 
have suggested that zero-temperature interfaces are actually
described by SLE \cite{slesg,fractalSG3d}.

For random ferromagnetic spin models,
one expects 'conventional scaling' as in Eq. \ref{enersgcriticonventional}
for the energy where $u_{E_c}$ is a random variable of order $O(1)$.
More generally, in the presence of relevant disorder, there is a lack of
self-averaging in all singular contributions of thermodynamic observables
in the sense that the leading term remains distributed
\cite{domany}. 
Note that for the Potts model with $q \geq 3$ states,
the interface becomes a non-directed branching object at criticality.
Some authors have studied the relevance of branching
within a solid-on-solid approximation 
where the 'directed' character of the low-temperature phase
is kept \cite{Kardar_branching,Cardy_branching}.
However, the 'directed' character is not expected to hold at criticality
for at least two reasons :
first, this 'directed'
character does not hold at criticality already for pure ferromagnets,
and second, in two dimensions, the directed polymer is always in its disordered
dominated phase, whereas ferromagnets undergo a phase transition 
where the disorder relevance of the Harris criterion depends on $q$
(see \cite{Cardy_branching} for a more detailed discussion).

The aim of this paper is to study numerically the critical behavior 
of some two-dimensional random Potts ferromagnet in the presence of relevant 
disorder.
We have chosen to work on the diamond hierarchical lattice
of effective dimension $d_{eff}=2$,
where large length scales can be studied via exact renormalization,
and with the Potts model with $q=8$ states 
so that disorder is relevant according to the Harris
criterion (see Appendix \ref{pottspure}). 
We present detailed results on the statistics of the interface free-energy,
energy, entropy and interfacial adsorption of non-boundary states.

\subsection{ Organization of the paper}

The paper is organized as follows.
In Section \ref{diamond}, we recall the exact renormalization equations
for the diamond hierarchical lattice, that are used to study numerically
the disordered Potts model with $q=8$ states 
on the diamond hierarchical lattice
of effective dimension $d_{eff}=2$.
We then describe our numerical results on 
 the interface free-energy statistics (Section \ref{secfree}),
on the interface energy and entropy statistics (Section \ref{secener}),
and on the interfacial adsorption of non-boundary states (Section
 \ref{nbstates}).
In Section \ref{sgdiamond}, we discuss the similarities and differences
with  the spin-glass transition in effective
dimension $d_{eff}=3$.
Finally we give our conclusions in Section \ref{conclusion}.
Appendix \ref{pottspure} contains a reminder on the pure Potts model
on hierarchical lattices.

\section{ Renormalization equations for spin models on hierarchical lattices } 

\label{diamond}

\subsection{ Reminder on the diamond hierarchical lattices}

\begin{figure}[htbp]
\includegraphics[height=6cm]{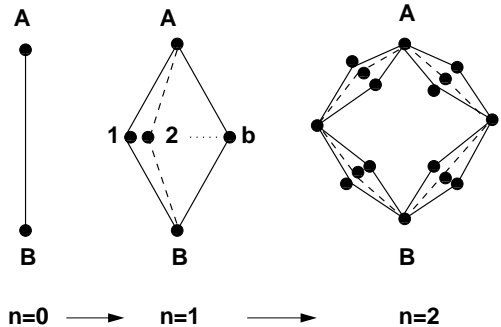}
\hspace{1cm}
\caption{ Hierarchical construction of the diamond lattice of
branching ratio $b$.   }
\label{figdiamond}
\end{figure}

Among real-space renormalization procedures \cite{realspaceRG}, 
Migdal-Kadanoff block renormalizations \cite{MKRG} play a special role
because they can be considered in two ways, 
 either as approximate renormalization procedures on hypercubic lattices,
or as exact renormalization procedures on certain hierarchical lattices
\cite{berker,hierarchical}.
One of the most studied hierarchical lattice is the
diamond lattice which is constructed recursively
from a single link called here generation $n=0$ (see Figure \ref{figdiamond}):
 generation $n=1$ consists of $b$ branches, each branch
 containing $2$ bonds in series ;
 generation $n=2$ is obtained by applying the same transformation
to each bond of the generation $n=1$.
At generation $n$, the length $L_n$ between the two extreme sites
$A$ and $B$ is $L_n=2^n$, and the total number of bonds is 
\begin{eqnarray}
B_n=(2b)^n = L_n^{d_{eff}(b)} \ \ \ {\rm \ \ with \  \ } 
d_{eff}(b)= \frac{ \ln (2b)}{\ln 2}
\label{deff}
\end{eqnarray}
where $d_{eff}(b)$ represents some effective dimensionality.

\subsection{ Spin models on hierarchical lattices }

On this diamond lattice, various
disordered models have been studied,
such as the diluted Ising model \cite{diluted}, 
ferromagnetic random Potts model \cite{Kin_Dom,Der_Potts,andelman} and
spin-glasses \cite{young,mckay,Gardnersg,bray_moore,muriel,thill}.
The random ferromagnetic Ising Hamiltonian reads
\begin{eqnarray}
H_{Ising}= - \sum_{<i,j>}  J_{i,j} S_i S_j
\label{defising}
\end{eqnarray}
where the spins take the values $S_i=\pm 1$
and where the couplings $J_{i,j}$ are positive random variables
(the spin-glass Hamiltonian corresponds to the case of couplings 
of random sign).
The random ferromagnetic Potts Hamiltonian is a generalization
where the variable $\sigma_i$ can take $q$ different values.
\begin{eqnarray}
H_{Potts}= - \sum_{<i,j>} 2 J_{i,j} \delta_{\sigma_i,\sigma_j}
\label{defpotts}
\end{eqnarray}
(We choose $(2J)$ to recover the Ising case for $q=2$)

\subsection{Renormalization equation for the interface free-energy  }

The free-energy cost $F^{inter}_L$ of creating an interface between
the two end-points $A$ and $B$
of the diamond lattice of Fig. \ref{figdiamond}
is defined by
\begin{eqnarray}
 F^{inter} \equiv  F_{+-} - F_{++}
\label{deffinterdiamond}
\end{eqnarray}
where $F_{++}= -T \ln Z_{++}$ and $F_{+-}=-T \ln Z_{+-}$ are the free-energies
corresponding respectively to the same color at both ends
or to two different colors at both ends. 
The renormalization equation are simpler to write in terms of the ratio
of the two partitions functions $Z_{++}$ and $Z_{+-}$
\begin{eqnarray}
x \equiv e^{ \beta F^{inter}}= \frac{Z_{++}}{Z_{+-}}
\label{defxrgpotts}
\end{eqnarray}
and one obtains \cite{Kin_Dom,Der_Potts,andelman}
\begin{eqnarray}
x_{n+1}= \prod_{i=1}^b \left( \frac{ x_n^{(i_1)} x_n^{(i_2)} + (q-1) }
{x_n^{(i_1)} + x_n^{(i_2)} + (q-2)} \right)
\label{rgpotts}
\end{eqnarray}

\subsection{ Renormalization equation for the interface energy }

The energy cost for creating an interface between
the two ends at distance $L$ is defined similarly by
\begin{eqnarray}
E^{inter}= E_{+-}-E_{++} 
= - \frac{d}{d \beta} \left( \ln Z_{+-}
- \ln Z_{++}  \right) = \frac{d}{d \beta} \ln \frac{Z_{++}}{Z_{+-}}
\label{defeinter}
\end{eqnarray}

The renormalization equation reads in terms of the variable 
$E \equiv E^{inter}$ and $x$ introduced above (Eq \ref{defxrgpotts})
\begin{eqnarray}
E_{n+1}= \sum_{i=1}^b
\frac{x^{(i_1)} (x^{(i_2)}-1)(x^{(i_2)}+q-1) E_{n}^{(i_1)}
+x^{(i_2)} (x^{(i_1)}-1)(x^{(i_1)}+q-1) E_{n}^{(i_2)} }
{(x^{(i_1)} + x^{(i_2)} +q-2) (x^{(i_1)}x^{(i_2)} +q-1)}
\label{rgener}
\end{eqnarray}

\subsection{ Renormalization equations for the order parameter
and the interfacial adsorption }

To study the order parameter and the interfacial adsorption,
let us introduce the notation 
\begin{eqnarray}
M^{a,b}_n= \sum_i < \delta_{\sigma_i,1} >_{a:b}
\label{defmab}
\end{eqnarray}
for the number of spins in state '1' on a hierarchical lattice at generation
$n$ when the two end points $A$ and $B$ of Fig. \ref{figdiamond}
are respectively in states 'a' and 'b'.
Using symmetries, one finally obtains closed renormalizations
for the following five variables 
\begin{eqnarray}
M^{(1,1)}_{n+1} && = \sum_{i=1}^b
\frac{ x_{n,i_1} x_{n,i_2} 
(M^{(1,1)}_{n,i_1} +M^{(1,1)}_{n,i_2} -1 )
+ (q-1)  (M^{(1,2)}_{n,i_1} +M^{(2,1)}_{n,i_2})}
{  x_{n,i_1} x_{n,i_2} 
+ q-1  }  \\
M^{(1,2)}_{n+1} && = \sum_{i=1}^b 
\frac{x_{n,i_1}   
(M^{(1,1)}_{n,i_1} +M^{(1,2)}_{n,i_2} -1)
+   x_{n,i_2} (M^{(1,2)}_{n,i_1} +M^{(2,2)}_{n,i_2})
+ (q-2)  (M^{(1,2)}_{n,i_1} +M^{(2,3)}_{n,i_2})  }
{ x_{n,i_1} 
+  x_{n,i_2} 
+ q-2  } \nonumber \\
 M^{(2,1)}_{n+1} && = \sum_{i=1}^b 
\frac{x_{n,i_2}  
(M^{(2,1)}_{n,i_1} +M^{(1,1)}_{n,i_2} -1)
+  x_{n,i_1} 
 (M^{(2,2)}_{n,i_1} +M^{(2,1)}_{n,i_2})
+ (q-2)  
(M^{(2,3)}_{n,i_1} +M^{(2,1)}_{n,i_2})  }
{ x_{n,i_2} 
+  x_{n,i_1} 
+ (q-2)  } \nonumber \\
M^{(2,2)}_{n+1} && = \sum_{i=1}^b 
\frac{ 
(M^{(2,1)}_{n,i_1} +M^{(1,2)}_{n,i_2} -1 )
+  x_{n,i_1} x_{n,i_2}(M^{(2,2)}_{n,i_1} +M^{(2,2)}_{n,i_2})
 + (q-2)  (M^{(2,3)}_{n,i_1} +M^{(2,3)}_{n,i_2}) }
{   x_{n,i_1} x_{n,i_2}
 + q-1 } \nonumber \\
M^{(2,3)}_{n+1} && = \sum_{i=1}^b 
\frac{ 
(M^{(2,1)}_{n,i_1} +M^{(1,2)}_{n,i_2} -1 )
+  x_{n,i_1} (M^{(2,2)}_{n,i_1} +M^{(2,3)}_{n,i_2}  )
+  x_{n,i_2} (M^{(2,3)}_{n,i_1} +M^{(2,2)}_{n,i_2}  )
 + (q-3)  (M^{(2,3)}_{n,i_1} +M^{(2,3)}_{n,i_2} ) } 
{  x_{n,i_1} +  x_{n,i_2} 
 + q-2  } \nonumber
\label{rgmfinal}
\end{eqnarray}
The order parameter can then be defined as
\begin{eqnarray}
M \equiv M^{(1,1)} - M^{(2,2)}
\label{deforder}
\end{eqnarray}
whereas the net absorption of non-boundary states of Eq. \ref{defnb} reads
\begin{eqnarray}
 N_{nb} \equiv  M^{(2,3)} - M^{(2,2)}
\label{defnbbis}
\end{eqnarray}

\subsection{ Numerical 'pool' method }

The numerical results presented below have been obtained
 with the so-called 'pool-method'
which is very often used for disordered systems on hierarchical lattices : 
the idea is to represent the probability distribution
$P_n(F_n,E_n)$ of the interface free-energy $F_n$ and energy $E_n$
at generation $n$, by a pool of $N$ realizations $\{(F_n^{(1)},E_n^{(1)}),..,
(F_n^{(N)},E_n^{(N)})) \}$.
The pool at generation $(n+1)$ is then obtained as follows :
each new realization $(F_{n+1}^{(i)},E_{n+1}^{(i)})$ is obtained by choosing 
$(2 b)$ realizations at random from the pool of generation $n$ and by applying
the renormalization equations given in Eq. \ref{rgpotts} and in Eq. \ref{rgener}.

The initial distribution of couplings was chosen to be 
\begin{eqnarray}
P_{Potts} (J) = \theta(J \geq 0) J e^{- \frac{J^2}{2} }
\label{pjpotts}
\end{eqnarray}
for the ferromagnetic Potts case, and 
Gaussian for the spin-glass case
\begin{eqnarray}
P_{SG} (J) = \frac{1}{\sqrt{2\pi} } e^{- \frac{J^2}{2} }
\label{gaussian}
\end{eqnarray}
At generation $n=0$ made of a single link (see Fig. \ref{figdiamond}), 
the free-energy and the energy of the interface coincide
and read in terms of the random coupling $J_i$ drawn with either
Eq \ref{gaussian} or Eq. \ref{pjpotts}
\begin{eqnarray}
F_{n=0}^{(i)}=E_{n=0}^{(i)}= 2 J_i 
\label{initial}
\end{eqnarray}
The numerical results presented below have been obtained with a pool of size
$N=4.10^7$ which is iterated up to $n=60$ or $n=80$ generations.
In the following sections, we study the random ferromagnetic
Potts $q=8$ on diamond lattice of effective dimension $d_{eff}=2$
corresponding to a branching ratio $b=2$.

\section{ Statistics of the interface free-energy  }

\label{secfree}

As recalled in the introduction, the interface free-energy
is expected to follow the scaling behavior of Eq \ref{thetadp}
below $T_c$ and to become a random variable of order $O(1)$ at $T_c$
(Eq \ref{freecriti}).
In this section, we present numerical results concerning the singularities
of the average and random contributions, 
as well as histograms, both in the low-temperature phase and at criticality.

\subsection{ Flow of the average value and width of the interface free-energy  }

\begin{figure}[htbp]
\includegraphics[height=6cm]{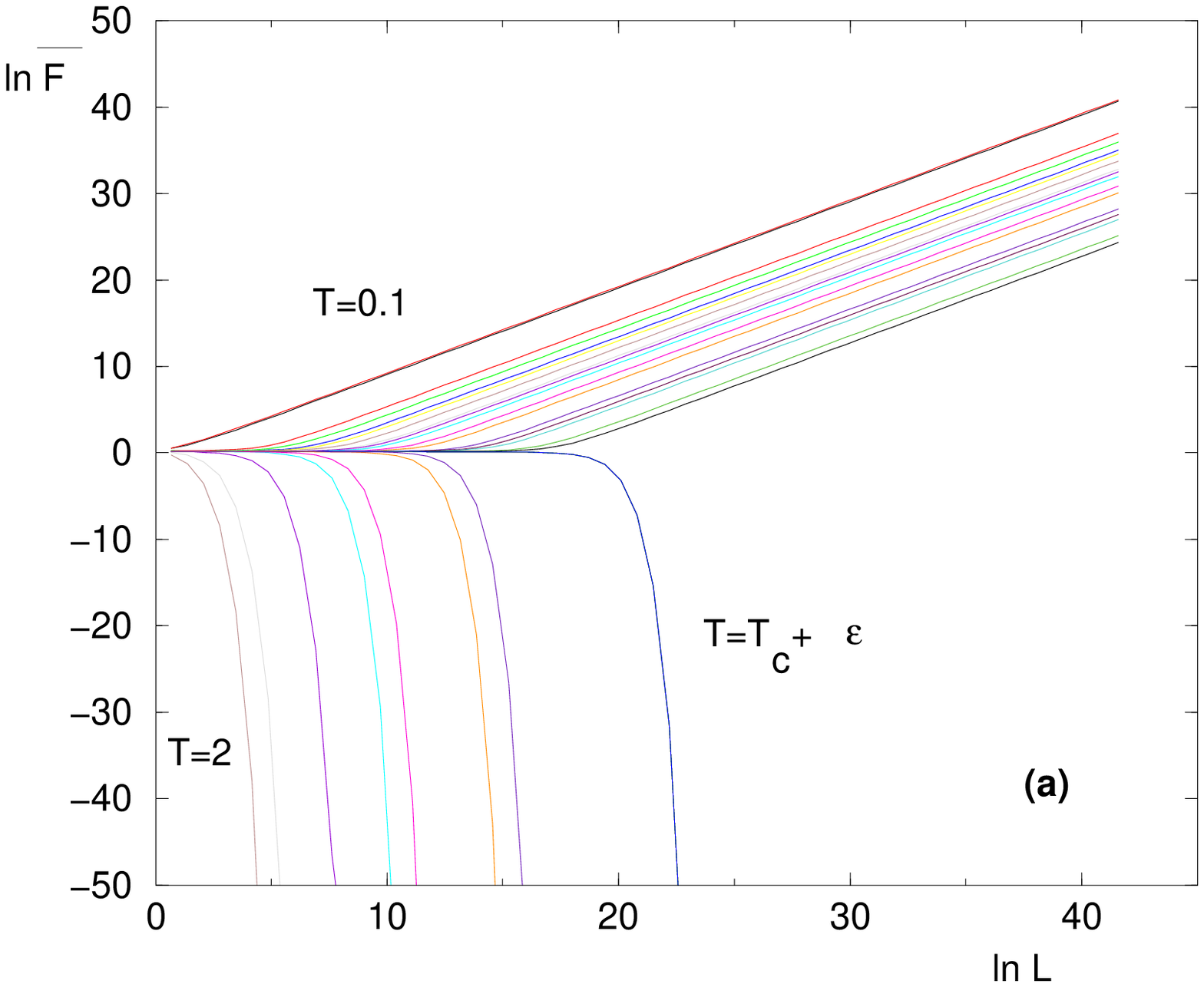}
\hspace{1cm}
\includegraphics[height=6cm]{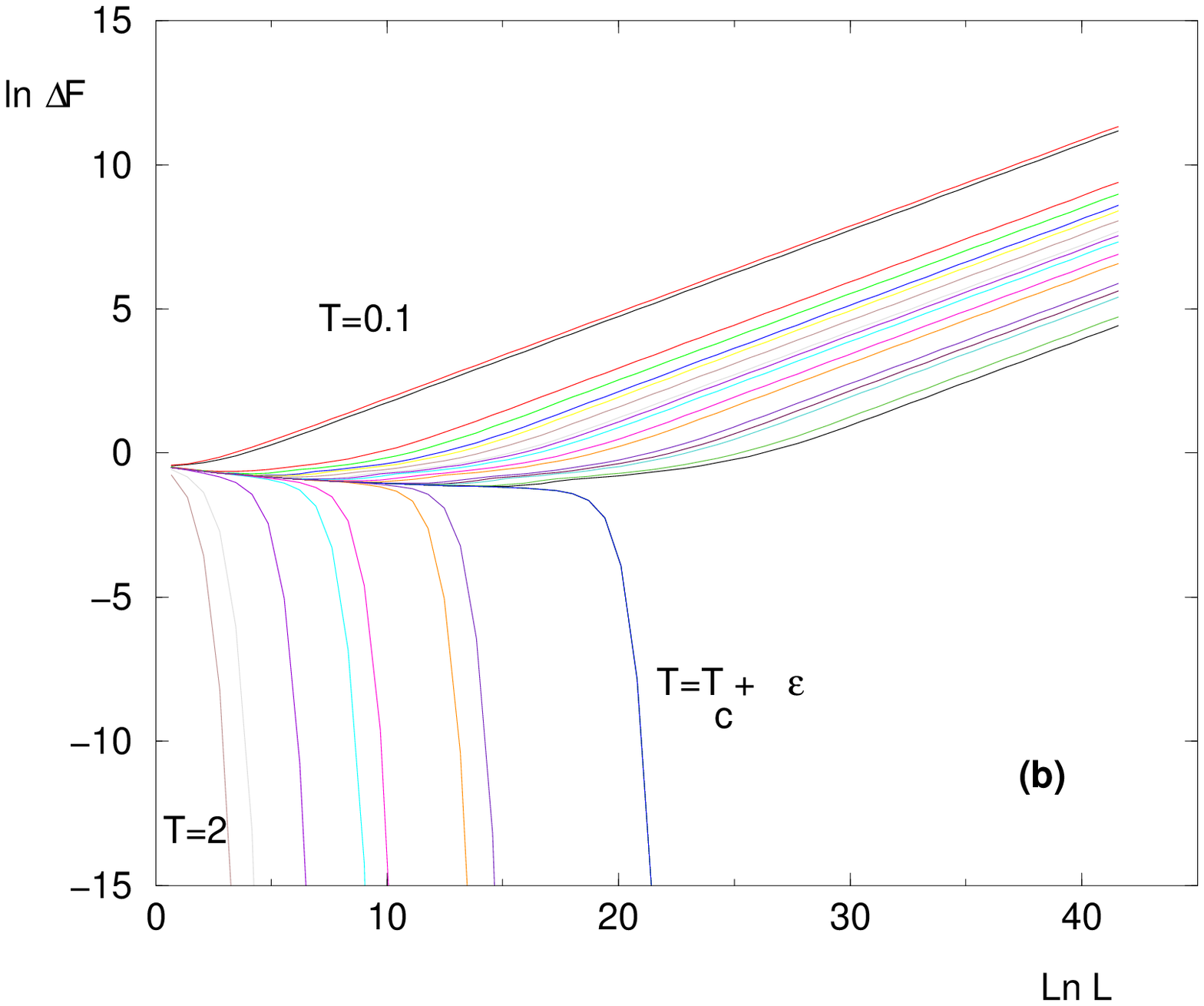}
\caption{(Color online) Flows of the average value and of the width
of the interface free-energy for many temperatures :
(a) log-log plot of the average value $\overline{ F}(L)$
of the free-energy distribution as a function of $L$.
(b) log-log plot of the width $\Delta F(L)$
of the free-energy distribution as a function of $L$.   }
\label{figpottsb2q8freewidth}
\end{figure}

The flows of the average free-energy $\overline{ F}(L)$ 
and of the free-energy width $\Delta F (L)$ 
are shown on Fig. \ref{figpottsb2q8freewidth} for many temperatures.
One clearly sees on these log-log plots
the two attractive fixed points separated by the critical temperature $T_c$.
The value of $T_c$ obtained by the pool method depends
on the pool, i.e. on the discrete sampling with $N$ values
of the continuous probability distribution. 
It is expected to converge towards the thermodynamic
critical temperature $T_c$ only in the limit $N \to \infty$.
Nevertheless, for each given pool,
the flow of free-energy allows a very precise
determination of this pool-dependent critical temperature,
for instance in the case considered 
$1.21685522 < T_c^{pool} < 1.21685523$.

For $T>T_c$, both the average free-energy 
and the free-energy width decay exponentially in $L$.
For $T<T_c$, the average free-energy grows asymptotically 
with the interface dimension $d_s=d_{eff}-1=1$ (see Eq \ref{thetadp})
\begin{eqnarray}
\overline{ F}(L) \simeq \left( \frac{L}{\xi_{av}(T)} \right)^{d_s} +...
\ \ \ {\rm with } \ \ d_s=1
\label{favbelowpottsq8}
\end{eqnarray}
where $\xi_{av}(T)$ is the correlation length
that diverges as $T \to T_c^-$.
The free-energy width grows asymptotically 
with the droplet exponent $\theta$ (see Eq \ref{thetadp})
\begin{eqnarray}
\Delta F(L) \simeq \left( \frac{L}{\xi_{var}(T)} \right)^{\theta(b)}
\ \ \ {\rm with } \ \ \theta(b=2) \simeq 0.299
\label{fwidthbelowpottsq8}
\end{eqnarray}
where $\xi_{var}(T)$ is the associated correlation length
that diverges as $T \to T_c^-$.
Note that $\theta(b=2) \simeq 0.299$ is the droplet exponent of the 
corresponding directed polymer model \cite{Der_Gri,diamondtails}.

 \subsection{ Divergence of the correlation lengths
 $\xi_{av}(T) $ and $\xi_{var}(T) $ } 

\begin{figure}[htbp]
\includegraphics[height=6cm]{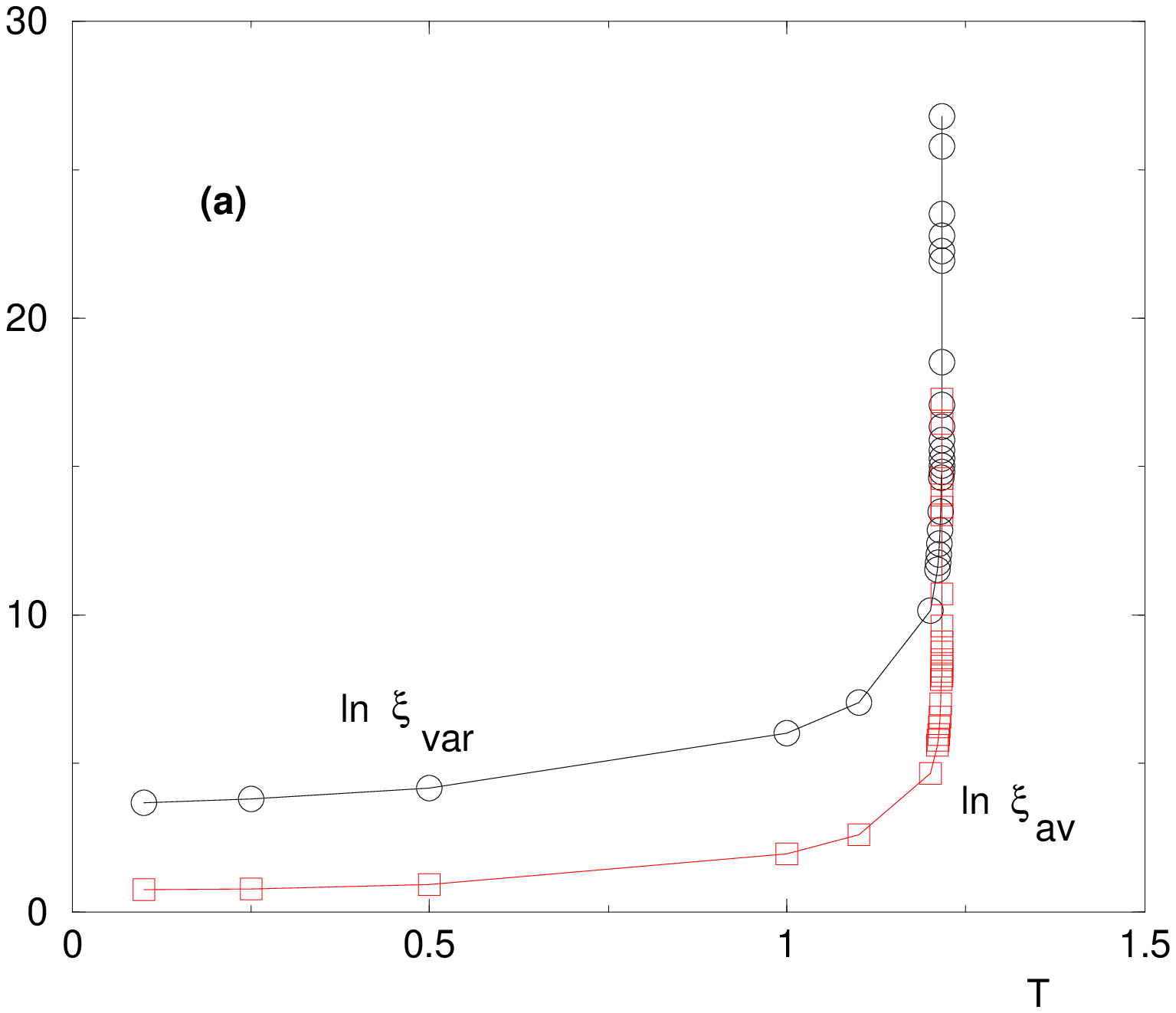}
\hspace{1cm}
\includegraphics[height=6cm]{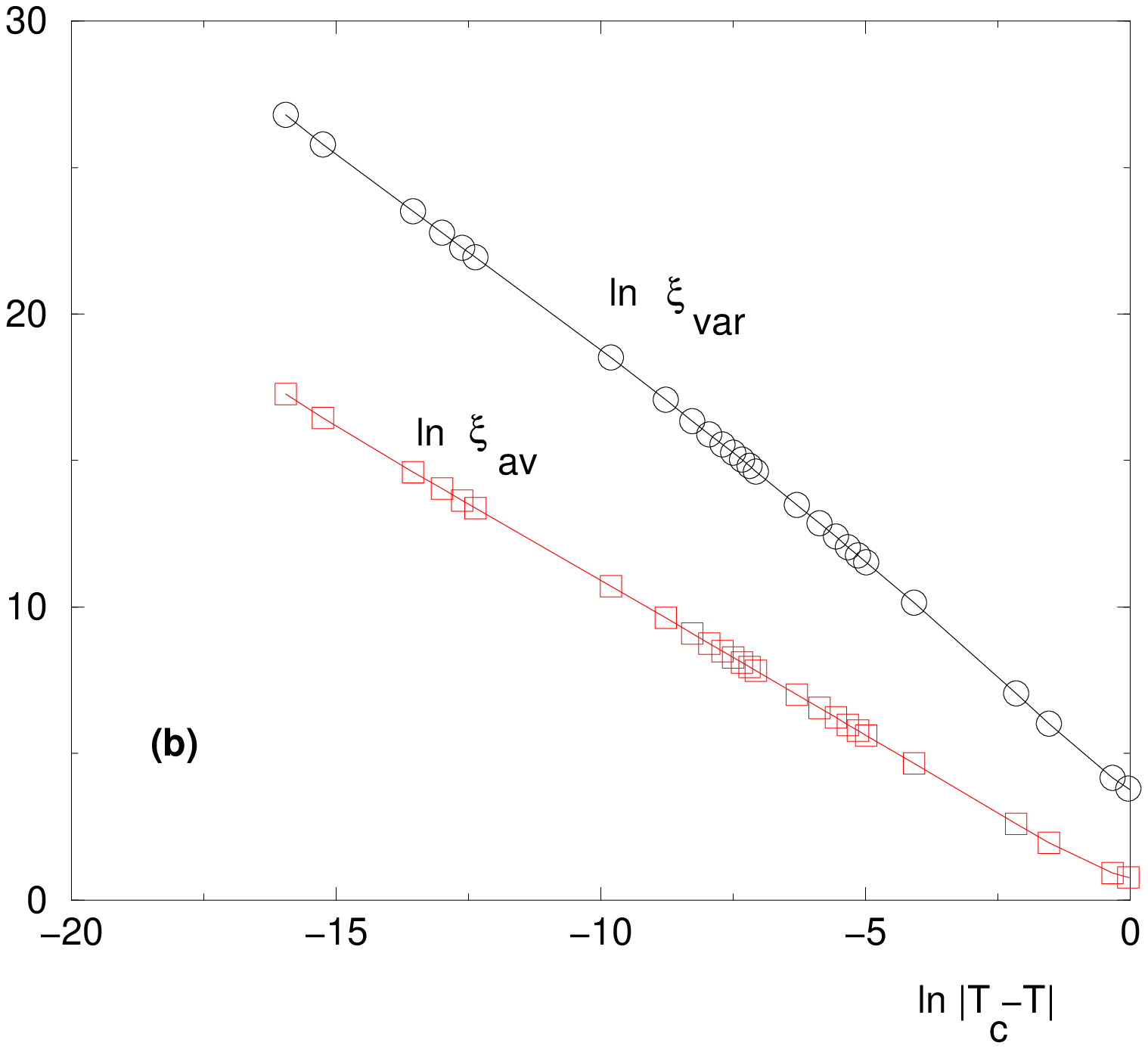}
\caption{ Divergence of the
correlation lengths $\xi_{av}(T)$ $(\square)$ and $\xi_{var}(T)$ $(\bigcirc)$ 
(a) $\ln \xi_{av}(T)$ and $\ln \xi_{var}(T)$ as a function of $T$
(b) $\ln \xi_{av}(T)$ and  $\ln \xi_{var}(T)$ as a function of 
$\ln \vert T_c-T \vert$  :
the asymptotic slopes are of order $\nu_{av} \simeq 1.07 $
and $\nu_{var} \simeq 1.34 $
}
\label{figlogxifreepottsb2q8}
\end{figure}

The correlation lengths  $\xi_{av}(T)$ and
$\xi_{var}(T)$ as measured from the free-energy average value 
(Eq \ref{favbelowpottsq8}) and from the free-energy
width (Eq \ref{fwidthbelowpottsq8} ) 
are shown on Fig. \ref{figlogxifreepottsb2q8} (a).
The log-log plot 
shown on Fig. \ref{figlogxifreepottsb2q8} (b) indicates power-law divergences
with two distinct correlation length exponents
\begin{eqnarray}
\xi_{av}(T) && \oppropto_{T \to T_c} (T_c-T)^{-\nu_{av}} 
\ \ { \rm with } \ \ \nu_{av} \simeq 1.07 \nonumber
\\
\xi_{var}(T) && \oppropto_{T \to T_c} (T_c-T)^{-\nu_{var}} 
\ \ { \rm with } \ \ \nu_{var} \simeq 1.34
\label{xifreenupottsq8}
\end{eqnarray}

In conclusion, our numerical results point towards the following
singular behavior for the interface free-energy (see Eq. \ref{thetadp})
\begin{eqnarray}
F^{inter}(L,T<T_c)  \oppropto_{T \to T_c^-} 
  \left( \frac{L}{\xi_{av}(T)} \right)^{d_s} 
+   \left( \frac{L}{\xi_{var}(T)} \right)^{\theta} u_F +...
\label{freedpsing}
\end{eqnarray}
where the average contribution and the random contribution 
involve two correlation lengths $\xi_{av}(T)$ and $\xi_{var}(T)$
that diverge with distinct exponents (Eq \ref{xifreenupottsq8}).
The presence of these two distinct correlation length exponents
in the interface free-energy
was a surprise for us, and we are not aware of any
 discussion of this possibility
in the literature.
The 'true' correlation length is expected to be $\xi_{av}(T)$
that appears in the extensive non-random contribution
to the interface free-energy. However, the presence of another 
length scale $\xi_{var}(T)$ that diverges with a larger exponent
remains to be better understood.

\subsection{ Histogram of the interface free-energy below $T_c$ }

\begin{figure}[htbp]
\includegraphics[height=6cm]{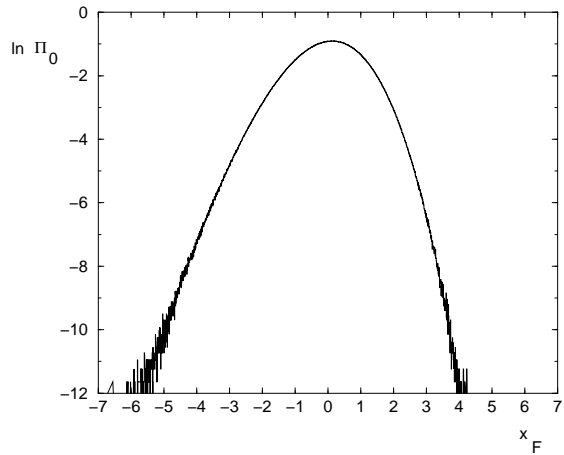}
\caption{ Statistics of the interface free-energy 
in the low-temperature phase (here $T=0.5$) : 
 Log representation of the distribution $\Pi_0(x_F)$ of 
the rescaled variable $x_F=\frac{F- \overline{F}}{\Delta F}$,
which present asymmetric tails (see Eq. \ref{pottstailsexponentslow}) }
\label{fighistot0.5potts8b2a}
\end{figure}

In the low-temperature phase, the 
 the interface free-energy is expected to 
follow the behavior of Eq \ref{thetadp},
where $u_F$ is a random behavior of order $O(1)$.
We show on Fig. \ref{fighistot0.5potts8b2a} 
the probability distribution $\Pi_0(x_F)$
of the rescaled variable $x_F=\frac{F- \overline{F}}{\Delta F}$
in log-scale to see the tails.
The two tails exponents $(\eta_-,\eta_+)$ defined by
\begin{eqnarray}
\ln \Pi_0(x_F) \opsimeq_{x_F \to \pm \infty} - \vert x_F \vert^{\eta_{\pm}}
\label{pottstailsfreelow}
\end{eqnarray}
are compatible with the relations proposed in our previous work
\cite{diamondtails} with $d_s=1$, $d=2$ and $\theta \simeq 0.299$ 
(see Eq \ref{fwidthbelowpottsq8})
\begin{eqnarray}
\eta_- && = \frac{d_s}{d_s-\theta} = \frac{1}{1-\theta} \sim 1.43
\nonumber  \\
\eta_+ && = \frac{d}{d_s-\theta} = \frac{2}{1-\theta} \sim 2.85
\label{pottstailsexponentslow}
\end{eqnarray}

\subsection{ Histogram of the interface free-energy at criticality}

\begin{figure}[htbp]
\includegraphics[height=6cm]{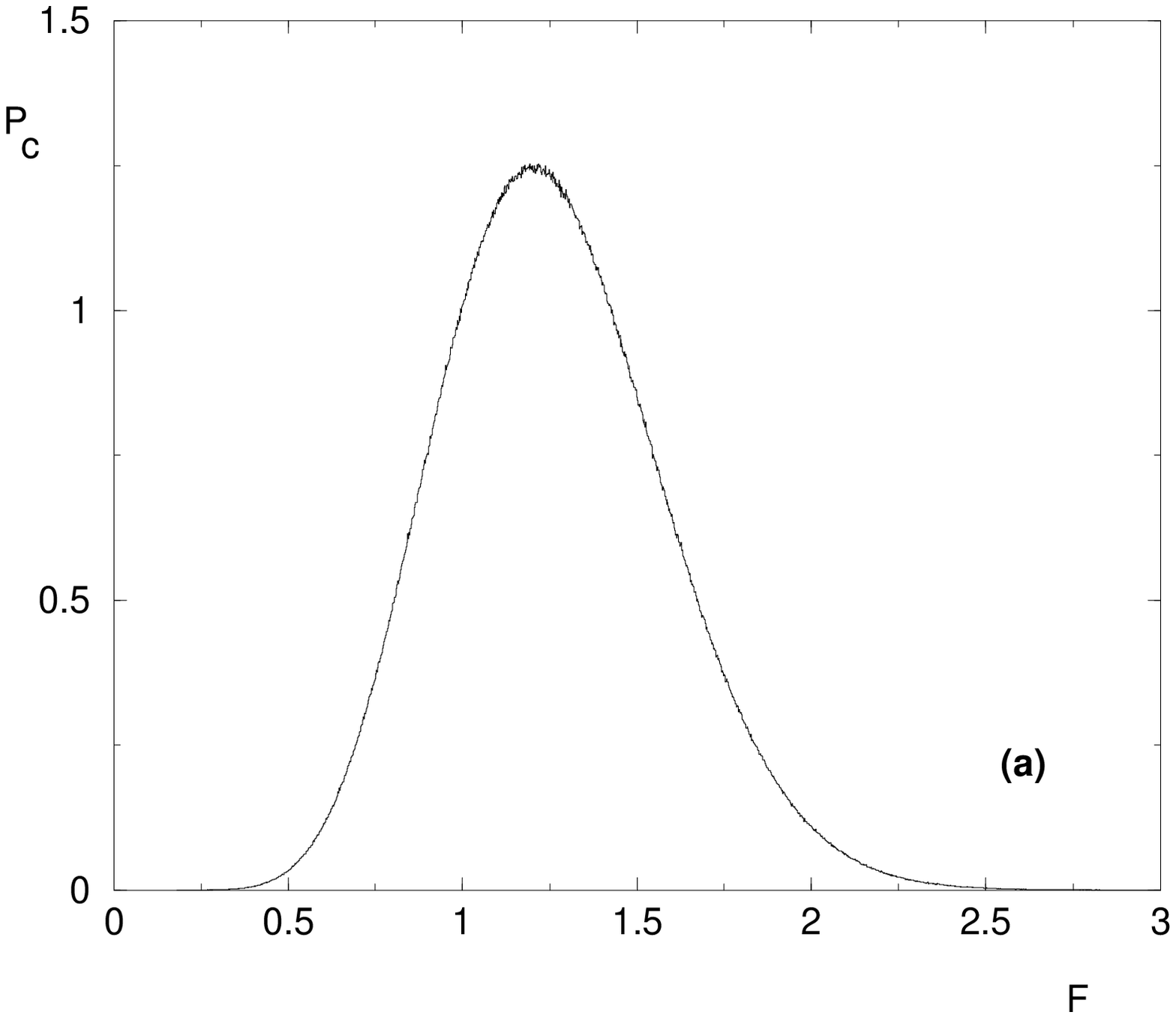}
\hspace{1cm}
\includegraphics[height=6cm]{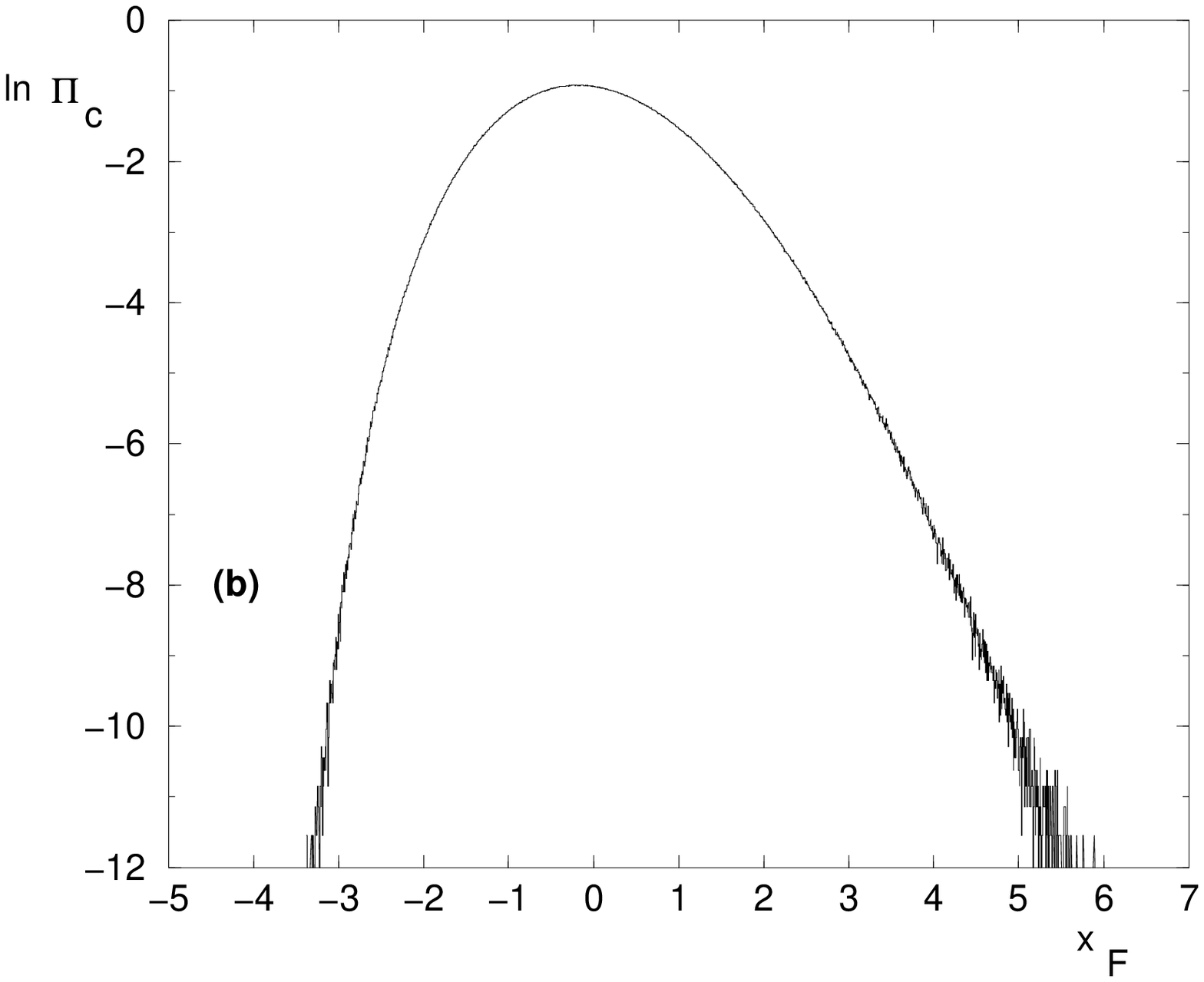}
\caption{ 
Histogram of the interface free-energy at criticality
(a) Unrescaled probability distribution $P_c(F)$ of the interface free-energy 
$F$ at $T_c$
(b) Log representation of the distribution $\Pi_c(x_F)$ of 
the rescaled variable $x_F=\frac{F- \overline{F}}{\Delta F}$
}
\label{fighistotcfreepotts8b2}
\end{figure}

At criticality, the interface free-energy is expected 
to become a random variable of order $O(1)$ (Eq. \ref{freecriti}) :
we show its probability distribution on Fig. \ref{fighistotcfreepotts8b2} (a).
To see the tails, we show in log-scale
the distribution of the rescaled variable
$x_F=\frac{F- \overline{F}}{\Delta F}$ 
on Fig. \ref{fighistotcfreepotts8b2} (b).

\section{ Statistics of the interface energy  }

\label{secener}

In this section, we present the numerical results concerning
the statistics of the interface energy.

\subsection{ Average and width of the interface energy  }

As recalled in the introduction, the interface energy
is expected to follow the scaling behavior of Eq \ref{enerdp}
below $T_c$. The extensive non-random part $e_0(T)$ is directly related
to the corresponding 
non-random part $f_0(T)$ of the free-energy of Eq. \ref{thetadp}
via the usual thermodynamic relation 
\begin{eqnarray}
e_0(T) = f_0(T)- T \ \frac{df_0(T)}{dT}
\label{thermoe0f0}
\end{eqnarray}
As a consequence, the singularity found previously for $f_0(T)$ 
(Eq \ref{favbelowpottsq8})
\begin{eqnarray}
f_0(T) \simeq \left( \frac{1}{\xi_{av}(T)} \right)^{d_s} 
\oppropto_{T \to T_c^-} (T_c-T)^{ d_s \nu_{av}} 
\label{singf0}
\end{eqnarray}
determines the singularity of $e_0(T)$ near $T_c$
\begin{eqnarray}
e_0(T) \oppropto_{T \to T_c^-} (T_c-T)^{ d_s \nu_{av}-1} 
\label{singe0}
\end{eqnarray}

\begin{figure}[htbp]
\includegraphics[height=6cm]{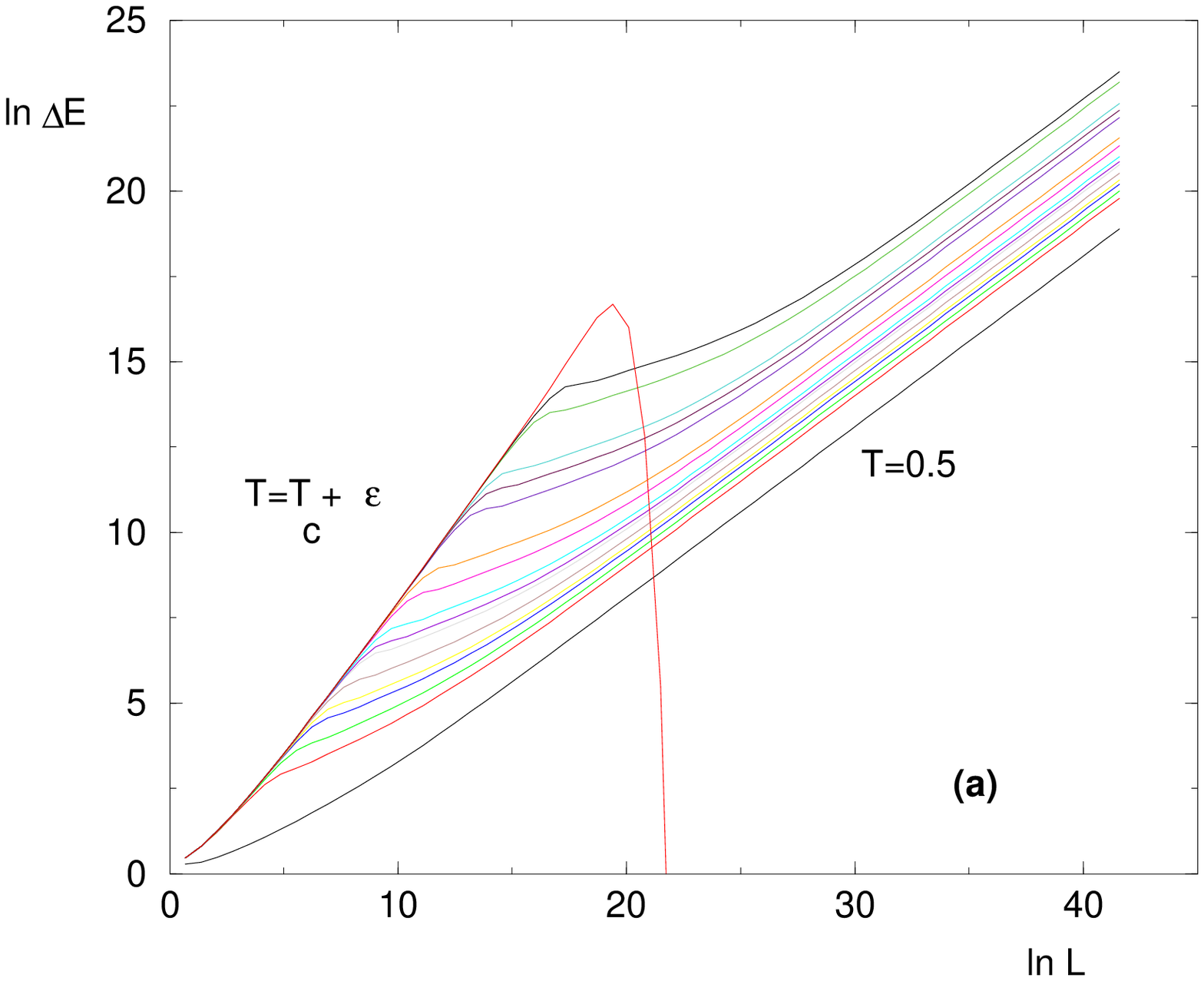}
\hspace{1cm}
 \includegraphics[height=6cm]{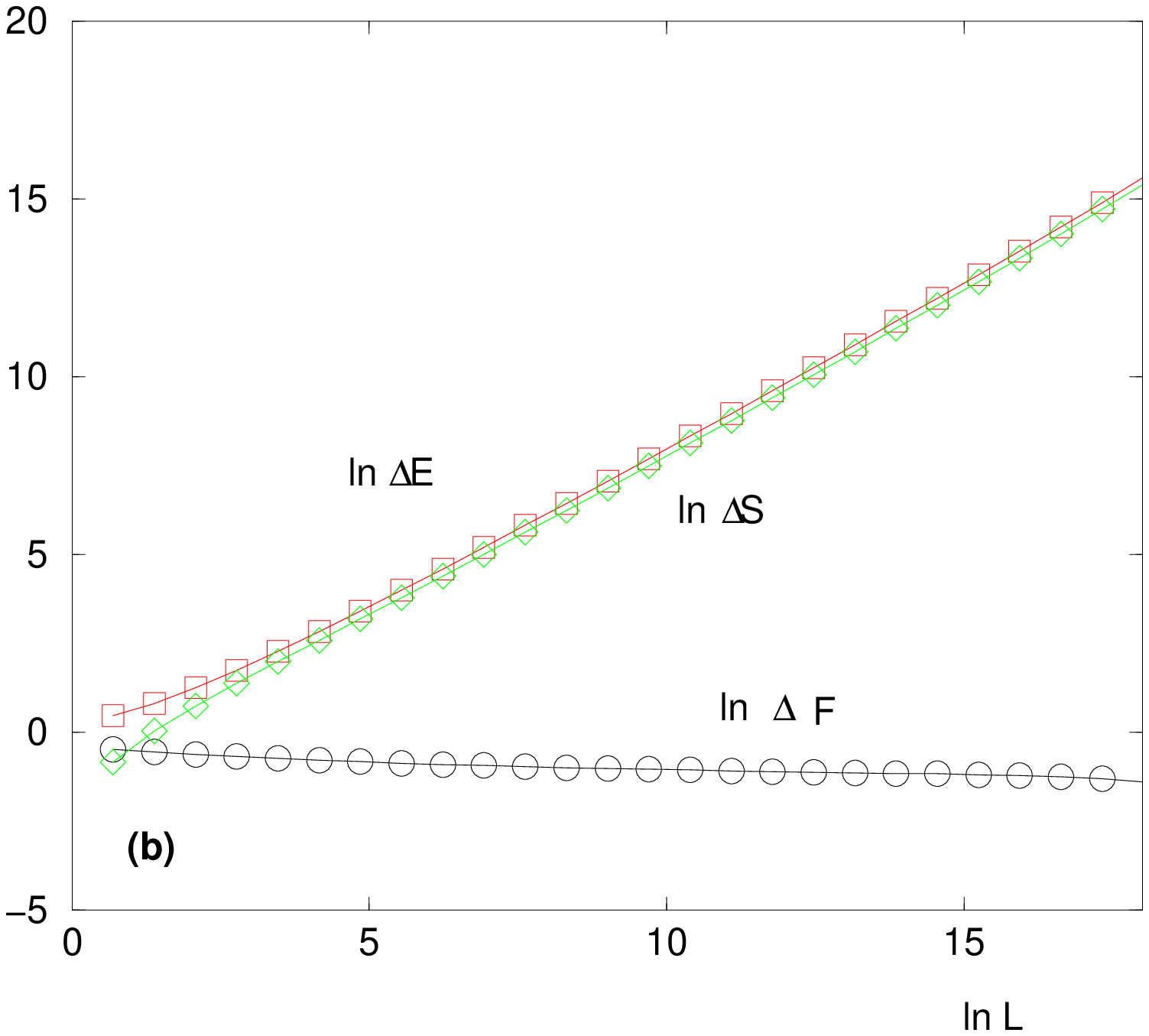}
\caption{(Color online) 
Flow of the width $\Delta E(L)$ 
of the energy distribution as $L$ grows
(a) $\ln \Delta E(L)$ as a function of $\ln L$ for many temperatures 
between $T=0.5$ and $T=T_c+\epsilon=1.21685523$
(b) Comparison
of $\ln \Delta E(L)$ ($\square$), $ \ln \Delta S(L)$ ($\Diamond$) and $ \ln \Delta F(L)$ ($\bigcirc$)
as a function of $\ln L$ at criticality.}
\label{figpottsb2q8enerwidth}
\end{figure}

We now consider the random contribution to the interface energy in Eq.
\ref{enerdp}. The flow of the width $\Delta E(L)$ as $L$ grows
is shown on Fig. \ref{figpottsb2q8enerwidth} for many temperatures.
For $T<T_c$, this width  grows asymptotically 
with the exponent $d_s/2=1/2$ as expected (see Eq \ref{enerdp})
\begin{eqnarray}
\Delta E(L) \simeq L^{\frac{d_s}{2}} = L^{\frac{1}{2}} 
\label{pottsq8ewidthbelow}
\end{eqnarray}

Exactly at criticality,
the energy width (and entropy width) grows as a power-law 
(see Fig \ref{figpottsb2q8enerwidth} b)
\begin{eqnarray}
\Delta E(L) \simeq L^{y_c} \ \ \ {\rm with } \ \ y_c \simeq 0.92
\label{pottsq8ewidthcriti}
\end{eqnarray}
This value for $y_c$ is in agreement with the value
 $1/\nu_{av} \simeq 0.93$ (see Eq. \ref{xifreenupottsq8}).

\begin{figure}[htbp]
\includegraphics[height=6cm]{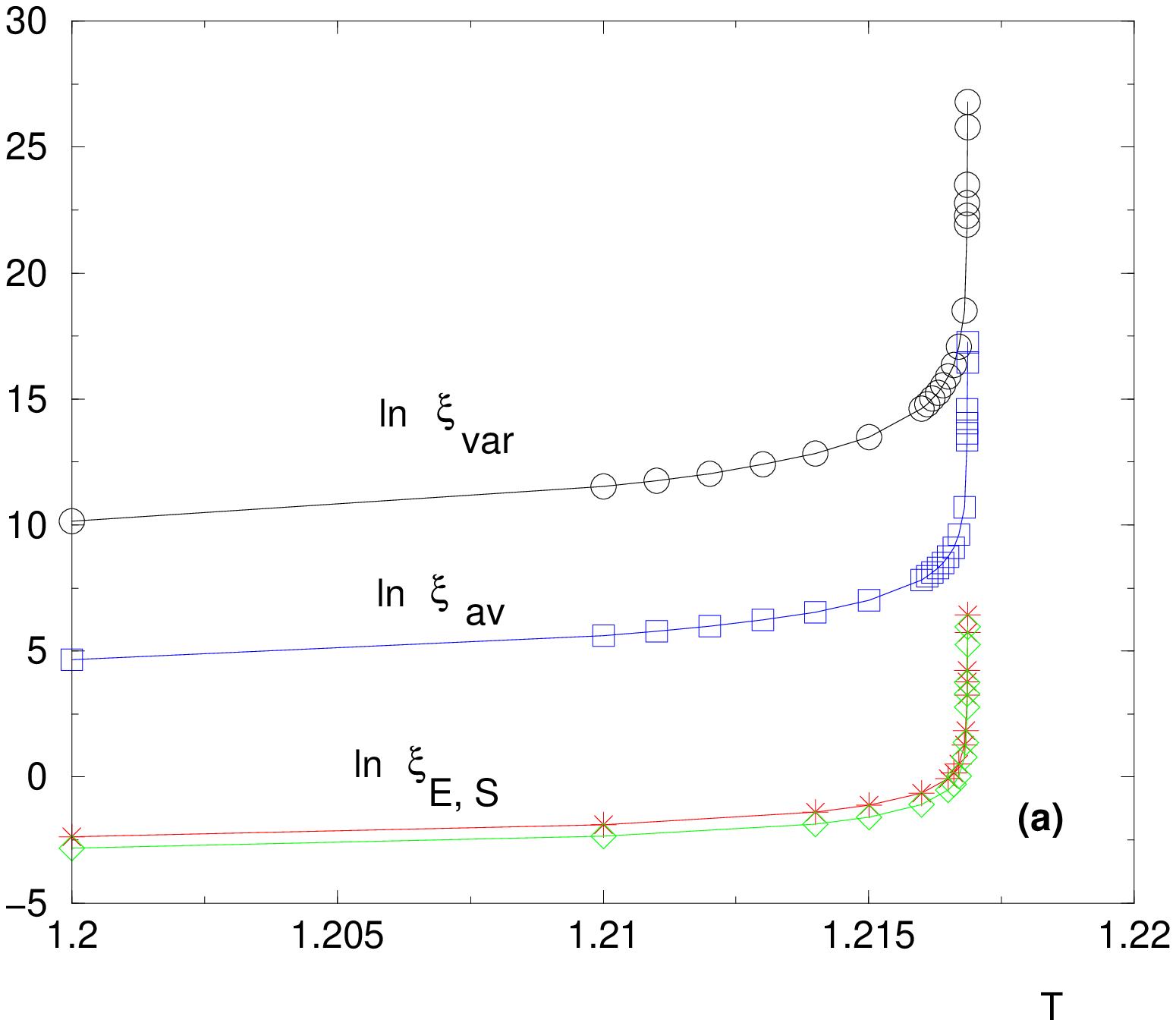}
\hspace{1cm}
\includegraphics[height=6cm]{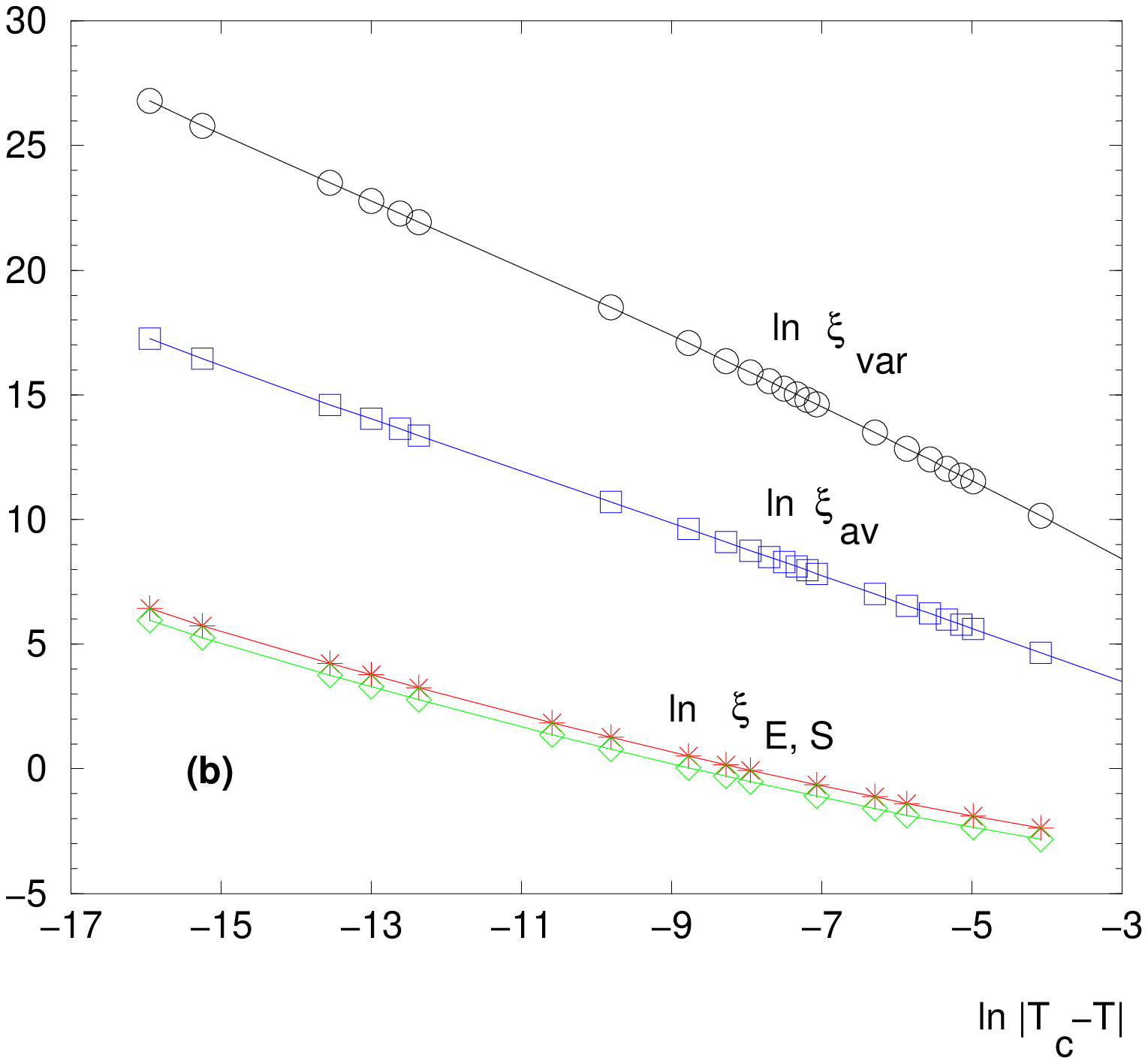}
\caption{(Color online)
Correlation lengths $\xi_E(T)$ ($\ast$) and $\xi_S(T)$ ($\Diamond$) as measured from the 
behavior of the energy and entropy widths 
(Eq. \ref{pottsq8ewidthfsslow})
(a) $\ln \xi_E(T)$ and $\ln \xi_S(T)$ as a function of $T$,
as compared to $\ln \xi_{av}(T)$ ($\square$) and $\ln \xi_{var}(T)$
($\bigcirc$) (b) $\ln \xi_E(T)$ and $\ln \xi_S(T)$ as a function of 
$\ln \vert T_c-T \vert$, as compared to $\ln \xi_{av}(T)$ and 
$\ln \xi_{var}(T)$
}
\label{figlogxienerentropiepotts8b2}
\end{figure}

We now define a correlation length 
$\xi_E(T)$ for $T<T_c$ via the finite-size scaling form
\begin{eqnarray}
\Delta E(L,T) \simeq L^{y_c} \Phi \left( \frac{L}{\xi_E(T)} \right)
\label{pottsq8ewidthfss}
\end{eqnarray}
In the regime $L \gg \xi_E(T)$, one should recover the $L$-dependence
of the low-temperature phase of Eq \ref{pottsq8ewidthbelow},
so the scaling function $\Phi(x)$ should present the asymptotic
behavior $\Phi(x) \sim x^{1/2-y_c} $
yielding the temperature dependence of the prefactor
\begin{eqnarray}
\Delta E(L,T) \opsimeq_{L \gg \xi_E(T)} 
\left( \frac{1}{\xi_E(T)} \right)^{1/2-y_c} L
\label{pottsq8ewidthfsslow}
\end{eqnarray}
One similarly may define a correlation length $\xi_S(T)$
from the finite-size scaling of the entropy width.

As shown on Fig \ref{figlogxienerentropiepotts8b2} b,
the log-log plot presents some curvature, 
so that the asymptotic slope $\nu_E$ defined by
\begin{eqnarray}
\xi_E(T) \opsimeq (T_c-T)^{-\nu_E} 
\label{pottsewidthxilow}
\end{eqnarray}
is difficult to measure precisely.
However, the slope $\nu_E$ is close to the value
$\nu_{av} \simeq 1.07$ of Eq. \ref{xifreenupottsq8}.

In conclusion, our numerical results point towards the following
singular behavior for the interface energy (see Eq. \ref{enerdp})
\begin{eqnarray}
E^{inter}(L,T<T_c)  \oppropto_{T \to T_c^-} 
\frac{1}{T_c-T} \left[  \left( \frac{L}{\xi_{av}(T)} \right)^{d_s} 
+   \left( \frac{L}{\xi_{av}(T)} \right)^{\frac{d_s}{2}} u_E +...\right]
\label{enerdpsing}
\end{eqnarray}
i.e. both the average contribution and the random contribution 
involve the same correlation length $\xi_{av}(T)$.
This result seems natural within the Fisher-Huse droplet theory
\cite{Fis_Hus,Fis_Hus_DP}
where the interface energy is a sum of random terms that follow
some Central Limit asymptotic behavior.
This picture is confirmed by the Gaussian distribution of the random
variable $u_E$ that we now consider.

\subsection{ Histogram of the interface energy below $T_c$ }

\begin{figure}[htbp]
\includegraphics[height=6cm]{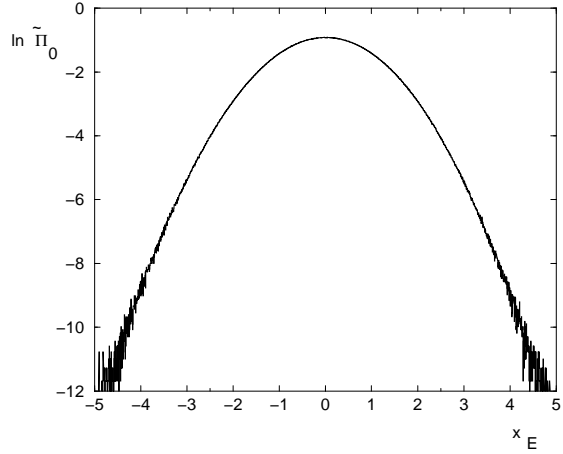}
\caption{
Statistics of the interface energy in the 
 low-temperature phase (here $T=0.5$) :
 Log representation of the distribution ${\tilde \Pi}_0(x_E)$ of 
the rescaled variable $x_E=\frac{E- \overline{E}}{\Delta E}$ :
it is a Gaussian}
\label{fighistot0.5potts8b2b}
\end{figure}

In the low-temperature phase, the 
 the interface energy is expected to 
follow the behavior of Eq \ref{enerdp},
where $u_E$ is a random variable of order $O(1)$
which is expected to be Gaussian distributed
within the droplet theory \cite{Fis_Hus,Fis_Hus_DP} :
this is in agreement with our numerical histogram of
the rescaled variable $x_E=\frac{E- \overline{E}}{\Delta E}$
shown on  Fig \ref{fighistot0.5potts8b2b}.

\subsection{ Histogram of the interface energy at criticality}

\begin{figure}[htbp]
\includegraphics[height=6cm]{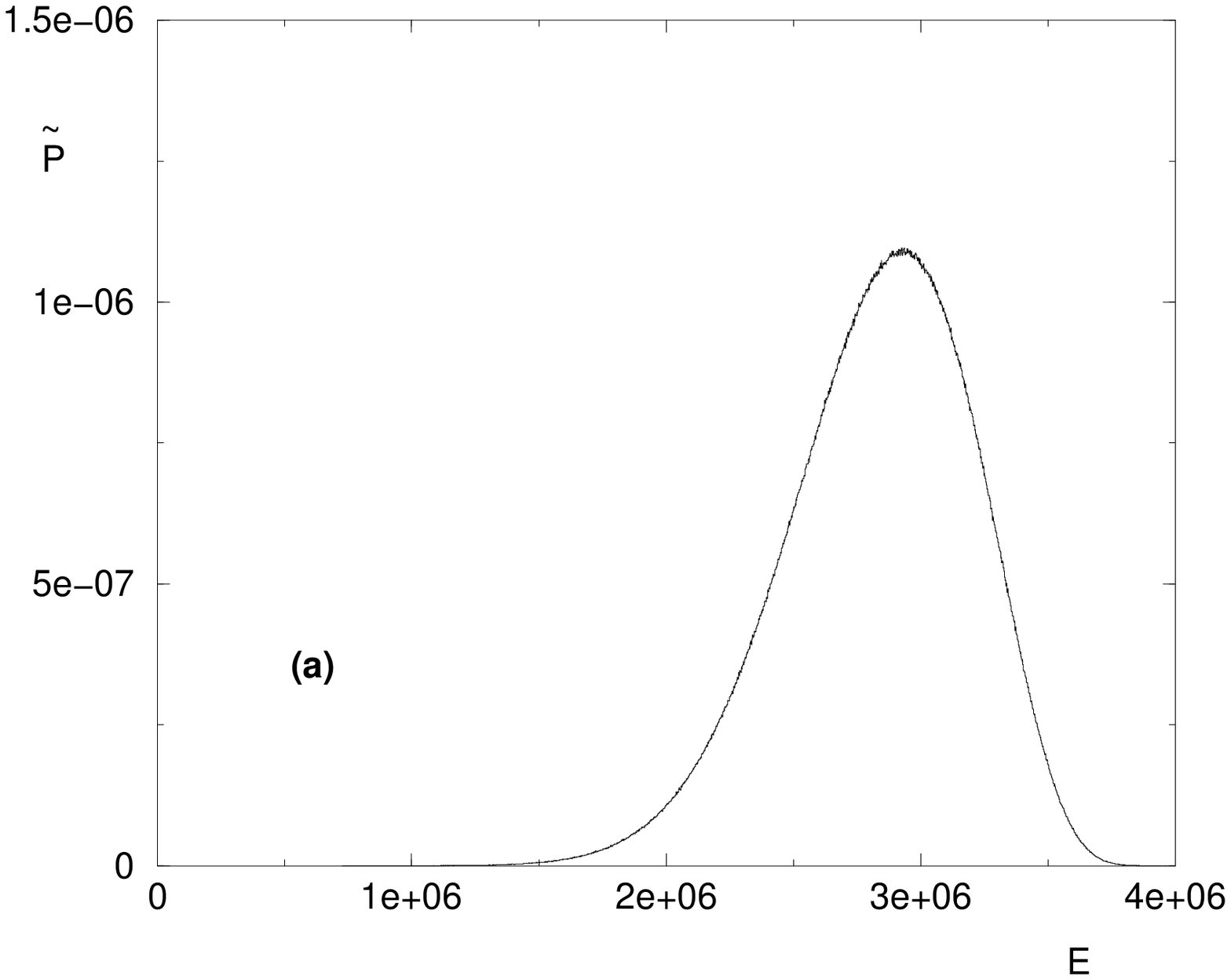}
\hspace{1cm}
\includegraphics[height=6cm]{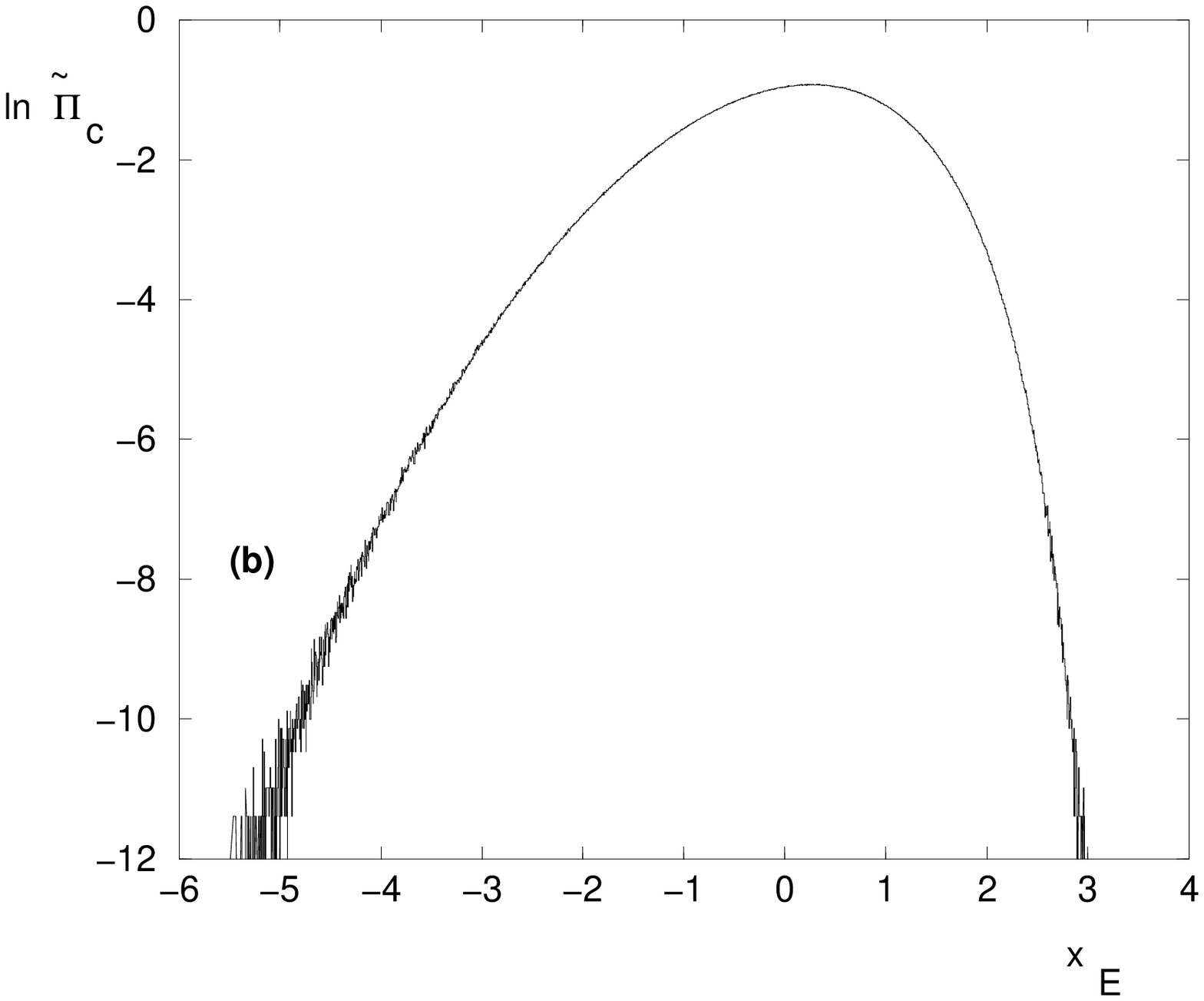}
\caption{ 
Histogram of the interface energy at criticality
(a) Unrescaled probability distribution $P_n(E)$ of the interface free-energy 
$E$ at $T_c$ at generation $n=22$
(b) Log representation of the distribution ${\tilde \Pi}_c(x_E)$ of 
the rescaled variable $x_E=\frac{E- \overline{E}}{\Delta E}$
}
\label{fighistotcenerpotts8b2}
\end{figure}

We show on Fig. \ref{fighistotcenerpotts8b2}
our numerical results for
the histogram of the interface energy at criticality :
the unrescaled distribution for $n=22$ generation
is shown on Fig. \ref{fighistotcenerpotts8b2} (a),
whereas the distribution of the rescaled variable
$x_E=\frac{E- \overline{E}}{\Delta E}$ is shown in log-scale
on Fig. \ref{fighistotcenerpotts8b2} (b).
The fast decay of the tails show that the scaling of the interface
energy
at criticality is well measured via its variance 
(see Eq. \ref{pottsq8ewidthcriti}).

\section{ Statistics of the order parameter and
of the interfacial adsorption }

\label{nbstates}

\subsection{ Statistics of the order parameter $M$ }

Below $T_c$, we find that the order parameter follows the scaling form
\begin{eqnarray}
 M(L,T<T_c) \simeq m_0(T) L^{d} + m_1(T) L^{\frac{d}{2}} u_i
\end{eqnarray}
The coefficient $m_0(T)$ of the extensive non-random term
vanishes at criticality with the exponent
\begin{eqnarray}
 m_0(T) \propto (T_c-T)^{\beta} \ \ {\rm with } \ \ \beta \simeq 0.163
\label{resbeta}
\end{eqnarray}
Exactly at criticality, the order parameter 
is expected to follow the behavior of Eq. \ref{nbtc},
up to a random variable $u_c$ of order $O(1)$
\begin{eqnarray}
 M(L,T_c) \simeq L^{d-\frac{\beta}{\nu}} u_c
\label{mtcdes}
\end{eqnarray}
Our previous measures of $\nu_{av} \simeq 1.07$ (Eq. \ref{xifreenupottsq8})
and $\beta \simeq 0.163$ (Eq. \ref{resbeta}) would correspond with $d=2$
to an exponent of order $d- \frac{\beta}{\nu_{av}} \simeq 1.85$.
 We measure (data not shown)
\begin{eqnarray}
 \overline{M}(L,T_c) \sim \Delta M(L,T_c) \sim L^{1.82} 
\label{measuremtcdes}
\end{eqnarray}
 in agreement with the scaling relation of Eq \ref{mtcdes}.

\subsection{ Statistics of the interfacial adsorption $N_{nb}$ }

\begin{figure}[htbp]
\includegraphics[height=6cm]{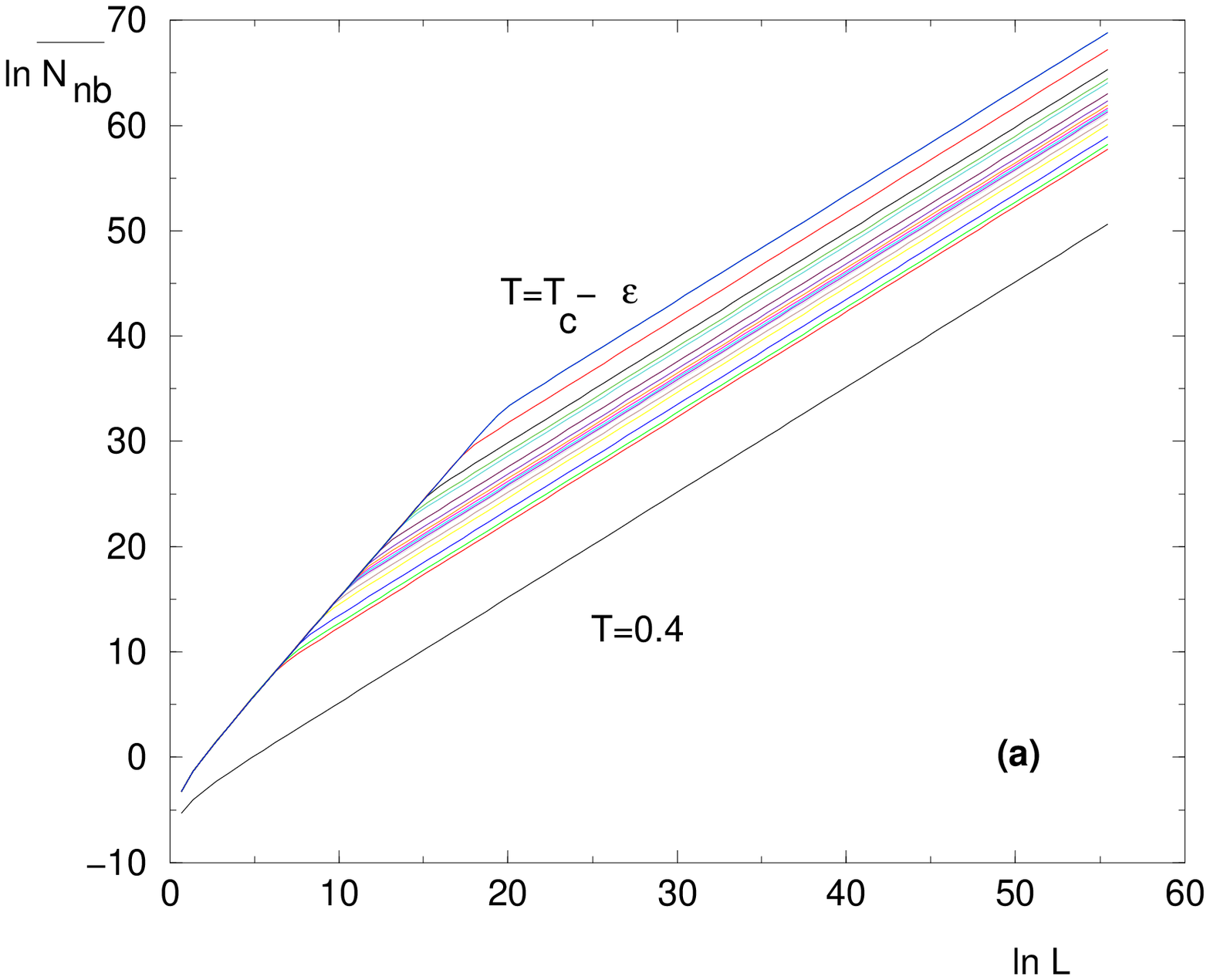}
\hspace{1cm}
\includegraphics[height=6cm]{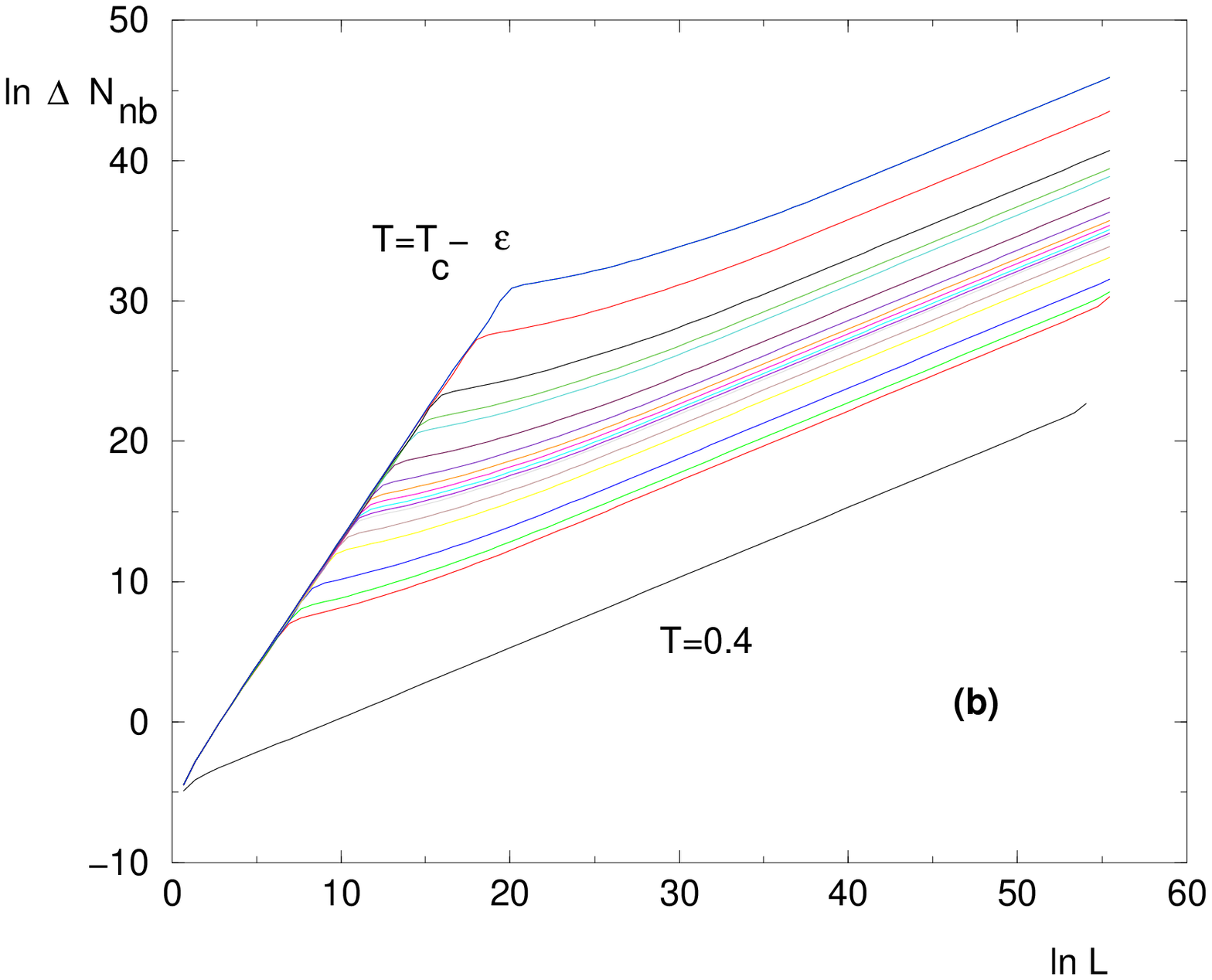}
\caption{(Color online) Flows of the average value and of the width
of the interfacial adsorption $N_{nb}$ for many temperatures :
(a) log-log plot of the average value $\overline{ N}_{nb}(L)$
 as a function of $L$.
(b) log-log plot of the width $\Delta N_{nb}(L)$ as a function of $L$.   }
\label{figpottsb2q8magneti}
\end{figure}

The flows of the average value and of the width of the net absorption
$N_{nb}$ of non-boundary states (defined in Eq \ref{defnb})
are shown on Fig. \ref{figpottsb2q8magneti}.
Below $T_c$, we find the scaling form
\begin{eqnarray}
 N_{nb}(T<T_c) \sim w_0(T) L^{d_s} + \delta(T) L^{\frac{d_s}{2}} v_i
\ \ {\rm with } \ \ d_s=1
\end{eqnarray}
The coefficient $w_0(T)$ of the extensive non-random term
is expected to diverge at criticality as in the pure case 
(Eq. \ref{singw0}).
The coefficient $\delta(T)$ of the random term
is expected to diverge to yield the same
finite-size scaling as Eq. \ref{nbtc}
exactly at criticality, so that
\begin{eqnarray}
 N_{nb}(L,T_c) \sim L^{d- \frac{\beta}{\nu}} v_{c}
\label{nbtcdes}
\end{eqnarray}
where $v_c$ is a random variable of order $O(1)$.
We measure (see Fig. \ref{figpottsb2q8magneti})
\begin{eqnarray}
 \overline{N}_{nb}(L,T_c) \sim \Delta {N}_{nb}(L,T_c) \sim L^{1.84} 
\label{measurenbtcdes}
\end{eqnarray}
again in agreement with the scaling relation of Eq \ref{nbtcdes}.

The conclusion of this section is that our numerical results concerning the order parameter
and the interfacial adsorption are consistent with 'conventional'
scaling in terms of the correlation length $\xi_{av}(T)$.

\section{ Comparison with the spin-glass
 on diamond lattice of effective dimension $d_{eff}=3$}

\label{sgdiamond}

The Migdal-Kadanoff renormalizations 
with a branching ratio $b=4$, which corresponds to an effective dimension
$d_{eff}=3$ (Eq \ref{deff}), have been much
used to study spin-glasses 
\cite{young,mckay,Gardnersg,bray_moore,muriel,thill}
(for the case $b=2$ corresponding to an effective dimension
$d_{eff}=2$ there is no spin-glass phase).
As recalled in the introduction, a new 'chaos exponent' $\zeta_c>1/\nu$
has been introduced in \cite{muriel,thill} to describe 
chaos properties at criticality, and this exponent was argued
to govern the energy at criticality (Eq \ref{enersgcritichaos}).
In the following, we confirm this scenario by directly measuring
the statistical properties of the interface energy.
We also discuss the similarities and
differences with the random ferromagnetic case
discussed above.

\subsection{ Flow of the width of the interface free-energy  }

\begin{figure}[htbp]
\includegraphics[height=6cm]{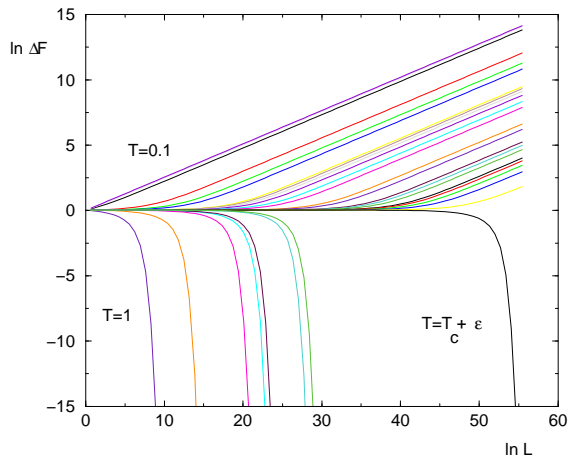}
\hspace{1cm}
\caption{(Color online) Spin-glass transition :
log-log plot of the width $\Delta F(L)$
of the free-energy distribution as a function of $L$, 
for many temperatures.   }
\label{figsgb4freewidth}
\end{figure}

In the spin-glass case, there is no non-random leading term 
(Eq \ref{thetasg}) in contrast to the random ferromagnetic case
( Eq \ref{thetadp}). As a consequence, we only have to consider here
the flow of the free-energy width $\Delta F(L)$ 
 shown on Fig. \ref{figsgb4freewidth}.
For $T>T_c$, the free-energy width decays exponentially in $L$.
For $T<T_c$, the free-energy width grows asymptotically 
with the droplet exponent $\theta$ (see Eq. \ref{thetasg})
\begin{eqnarray}
\Delta F(L) \simeq \left( \frac{L}{\xi_{var}(T)} \right)^{\theta(b)}
\ \ \ {\rm with } \ \ \theta(b=4) \simeq 0.255
\label{fwidthbelowsg}
\end{eqnarray}
where $\xi_{var}(T)$ is the correlation length
that diverges as $T \to T_c^-$.
The exponent $\theta(b=4) \simeq 0.255$ is in agreement with previous measures 
\cite{young,bray_moore,muriel}.

Again, the critical temperature obtained by this pool method depends
on the pool, i.e. on the discrete sampling with $N$ values
of the continuous probability distribution. 
It is expected to converge towards the thermodynamic
critical temperature $T_c$ only in the limit $N \to \infty$.
Nevertheless, for each given pool,
the flow of free-energy width allows a very precise
determination of this pool-dependent critical temperature,
for instance in the case considered 
$0.8810237 < T_c^{pool} < 0.88102375$.
This value is in agreement with previous measures
for a Gaussian initial condition  
\cite{young,bray_moore,muriel}.

\begin{figure}[htbp]
\includegraphics[height=6cm]{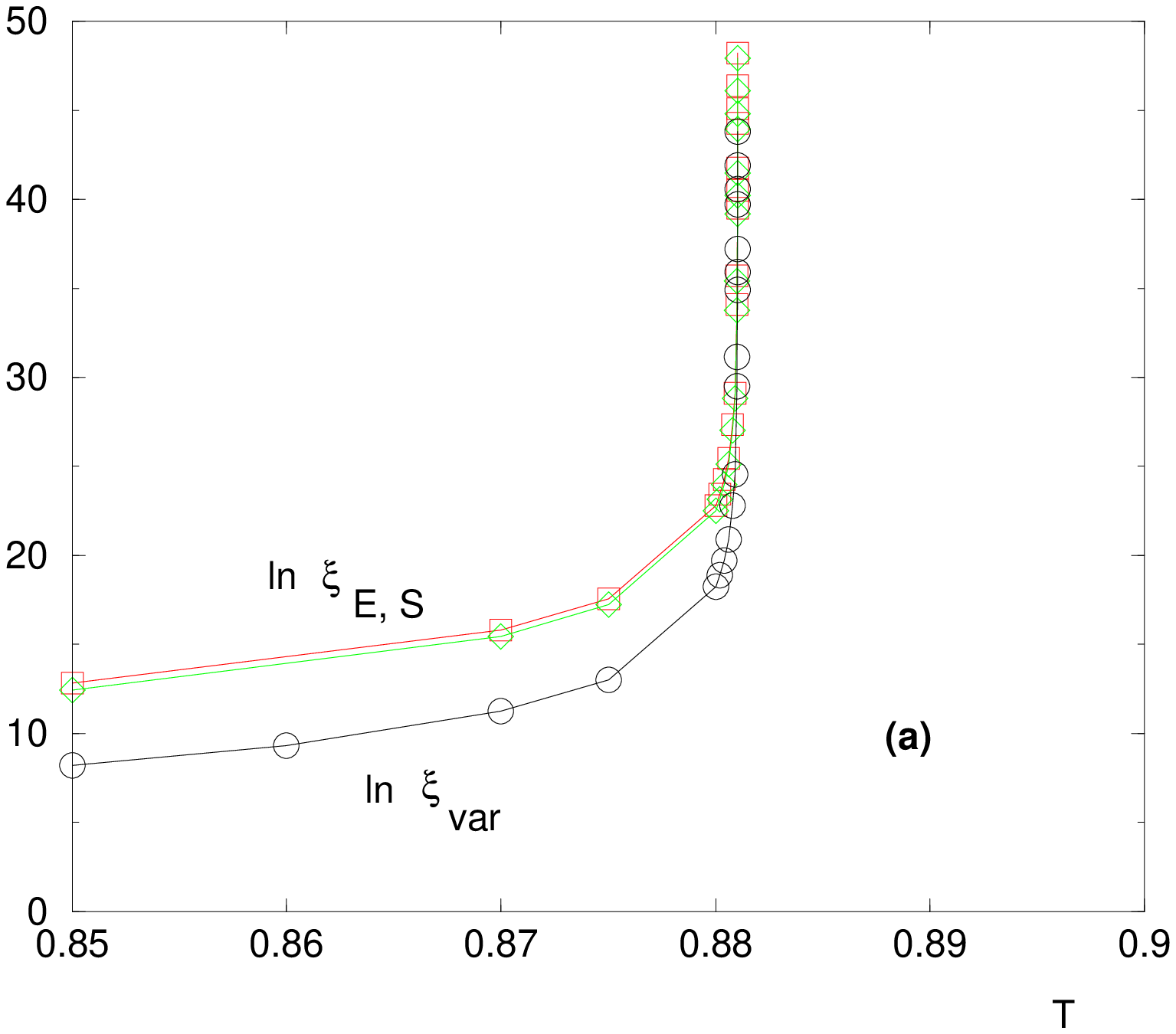}
\hspace{1cm}
\includegraphics[height=6cm]{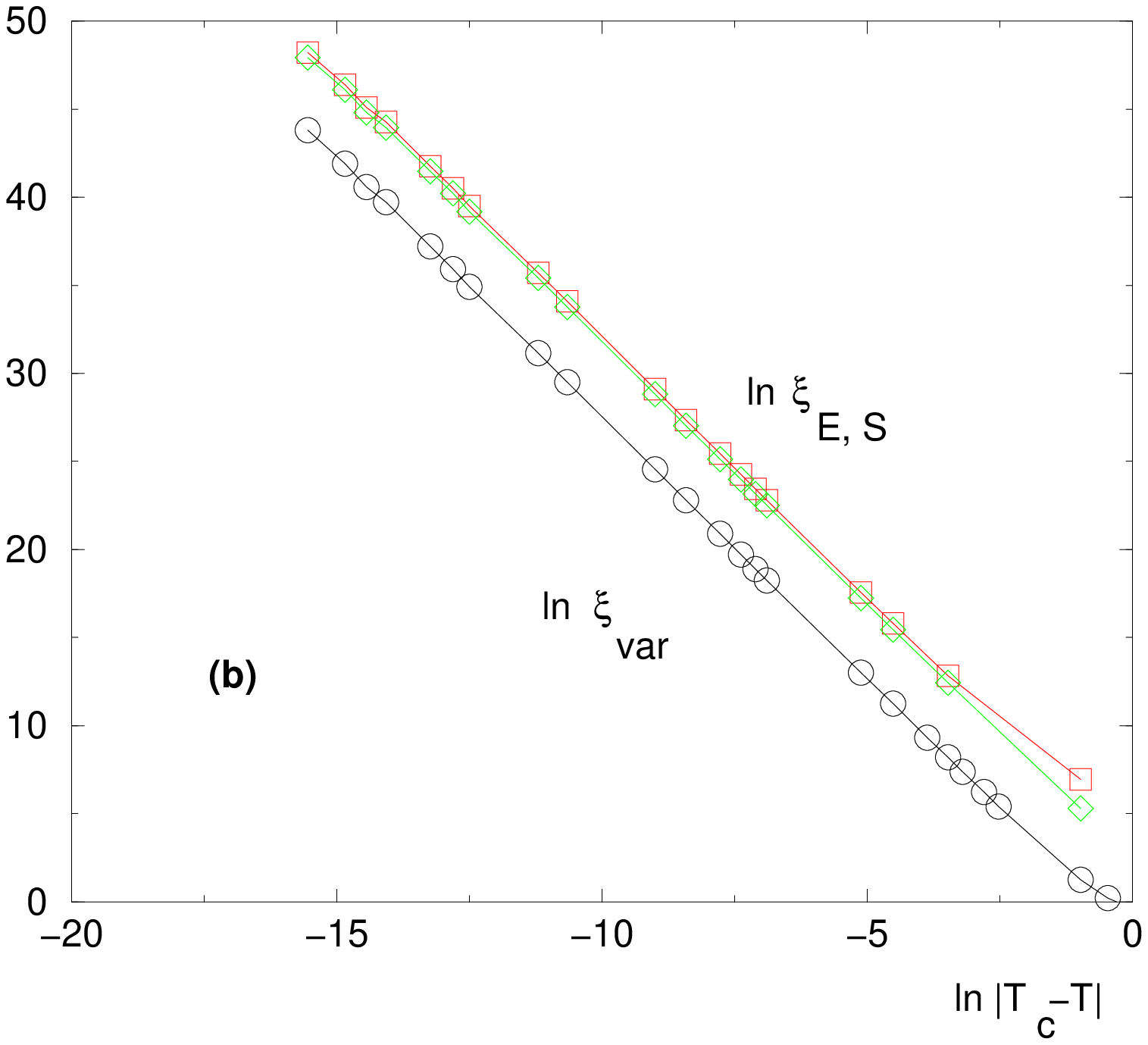}
\caption{(Color online) Spin-glass transition :
Correlation lengths $\xi_{var}(T)$ ($\bigcirc$), $\xi_E(T)$
($\square$) and $\xi_S(T)$ ($\Diamond$)
as measured from the 
behavior of the widths of the free-energy, energy and entropy 
distributions
(a) $\ln \xi_{var}(T)$, $\ln \xi_E(T)$ and $\ln \xi_S(T)$ as a function of $T$
(b) $\ln \xi_{var}(T)$, $\ln \xi_E(T)$ and $\ln \xi_S(T)$ as a function of 
$\ln \vert T_c-T \vert$,  :
the asymptotic slope is the same and of order $\nu_{var} \simeq 2.92 $
}
\label{figlogxienerentropiesgb4}
\end{figure}

The correlation length $\xi_{var}(T)$ as measured from the free-energy
width asymptotic behavior below $T_c$ (Eq \ref{fwidthbelowsg} ) 
is shown on Fig. \ref{figlogxienerentropiesgb4} (a).
The plot in terms of the variable $\ln (T_c-T)$
shown on Fig. \ref{figlogxienerentropiesgb4}
 (b) indicates a power-law divergence
\begin{eqnarray}
\xi_{var}(T) \oppropto_{T \to T_c} (T_c-T)^{-\nu_{var}} 
\ \ { \rm with } \ \ \nu_{var} \simeq 2.92
\label{xifreenusg}
\end{eqnarray}

\subsection{ Flow of the width of the interface energy } 

\begin{figure}[htbp]
\includegraphics[height=6cm]{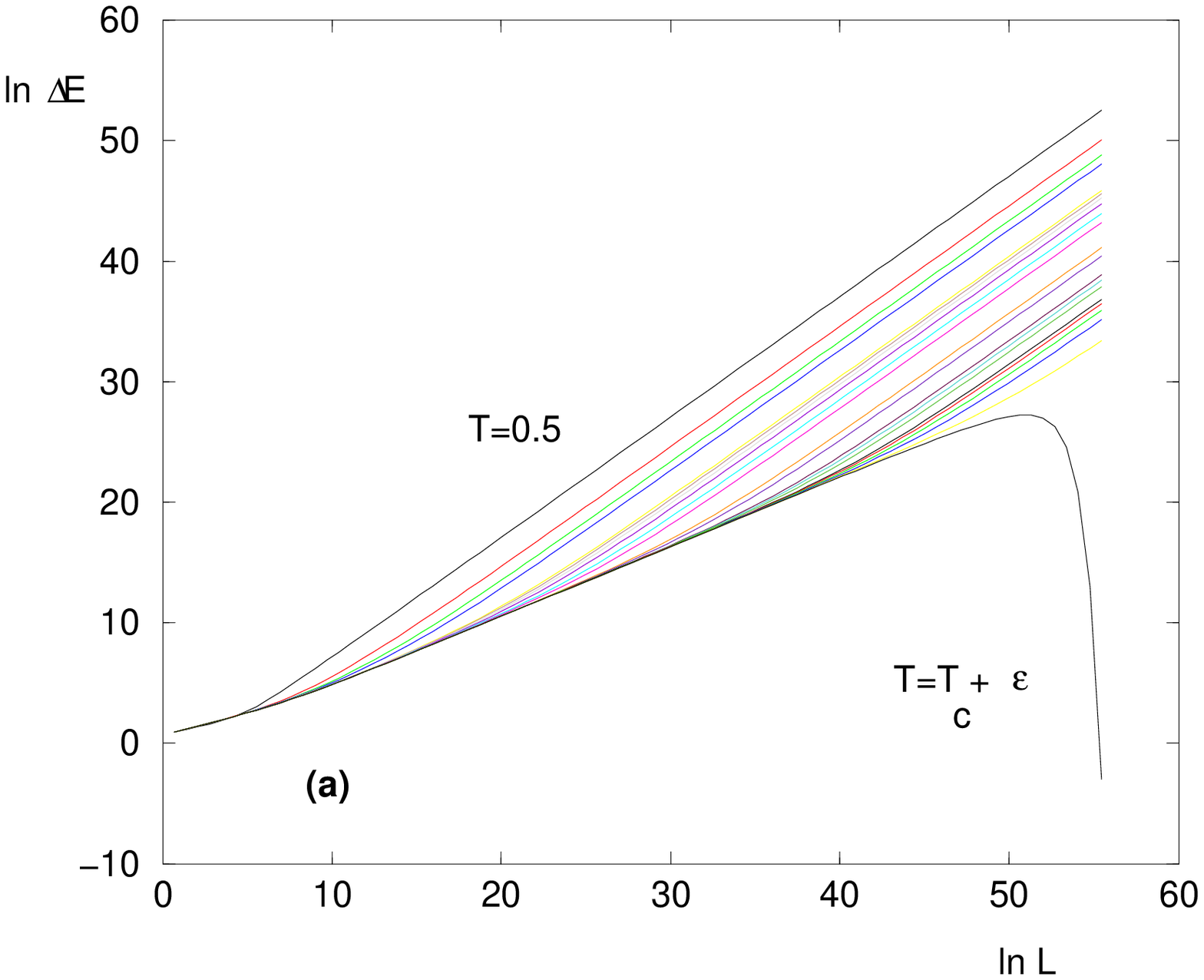}
\hspace{1cm}
 \includegraphics[height=6cm]{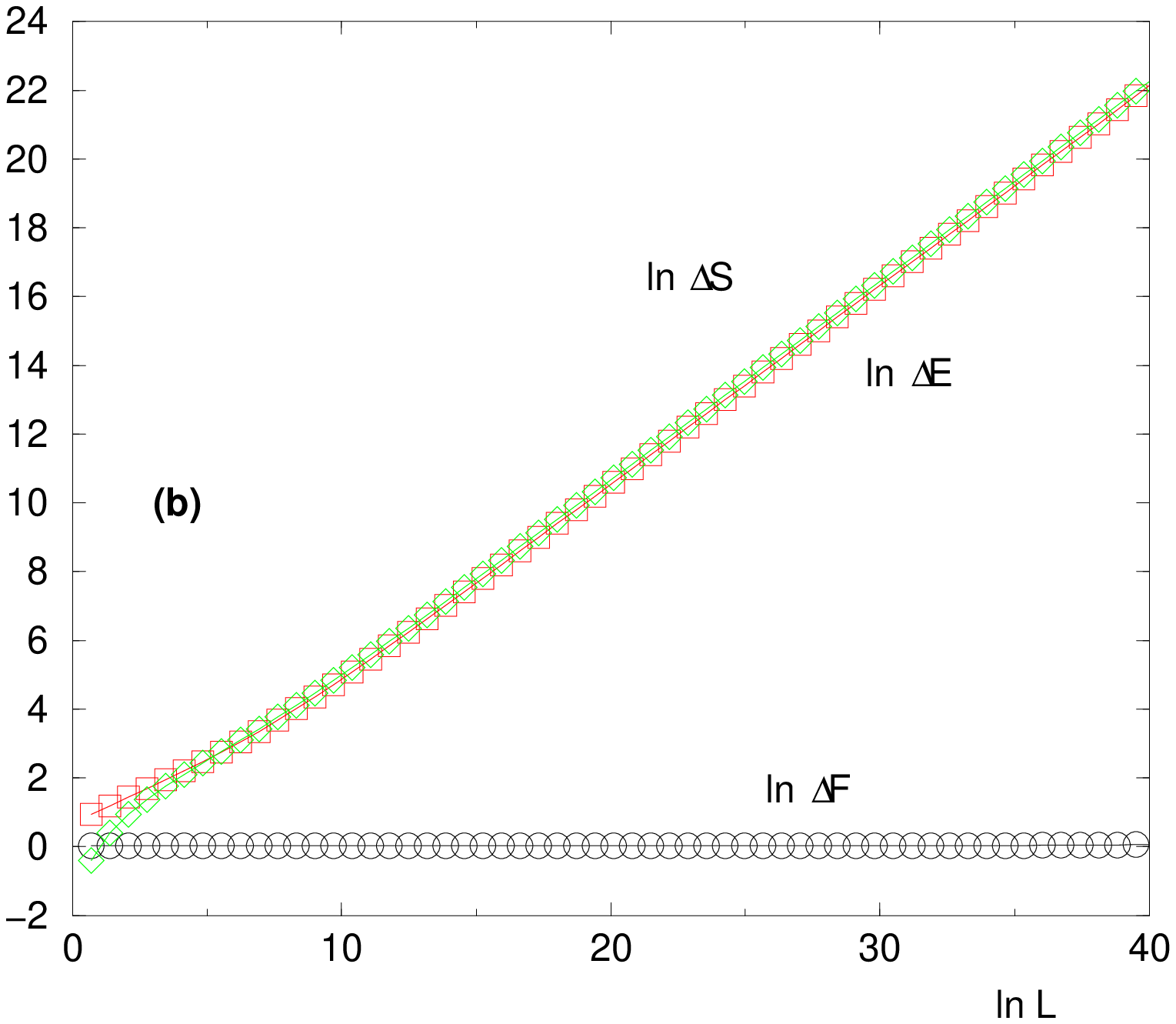}
\caption{(Color online) Spin-glass transition :
Flow of the width $\Delta E(L)$ 
of the energy distribution as $L$ grows
(a) $\ln \Delta E(L)$ as a function of $\ln L$ for many temperatures 
between $T=0.5$ and $T=T_c+\epsilon=0.88102375$
(b) Comparison
of $\ln \Delta E(L)$ , $ \ln \Delta S(L)$ and $ \ln \Delta F(L)$
as a function of $\ln L$ at criticality ($T_c^{pool}=0.8810237$).}
\label{figsgb4enerwidth}
\end{figure}

The flow of the energy width $\Delta E(L)$ 
as $L$ grows
are shown on Fig. \ref{figsgb4enerwidth} for many temperatures.
For $T<T_c$, 
this width  grows asymptotically 
with the exponent $d_s/2$ (see Eq. \ref{enersg})
\begin{eqnarray}
\Delta E(L) \simeq L^{\frac{d_s}{2}} \simeq L \ \ \ {\rm with } \ \ 
d_s(b=4)=d_{eff}(b=4)-1=2
\label{sgewidthbelow}
\end{eqnarray}

Exactly at criticality,
 the energy width (and entropy width) grows as a power-law 
(see Fig \ref{figsgb4enerwidth} b)
\begin{eqnarray}
\Delta E(L,T_c) \simeq L^{\zeta_c} \ \ \ {\rm with } 
\ \ \zeta_c \simeq 0.58
\label{sgewidthcriti}
\end{eqnarray}
This exponent is clearly greater than 
 $1/\nu_{var} \simeq 0.34$ (see Eqs \ref{xifreenusg})
and coincides with the chaos critical exponent measured in \cite{muriel}.

We now define a correlation length 
$\xi_E(T)$ for $T<T_c$ via the finite-size scaling form
\begin{eqnarray}
\Delta E(L,T) \simeq L^{\zeta_c} \Phi \left( \frac{L}{\xi_E(T)} \right)
\label{sgewidthfss}
\end{eqnarray}
In the regime $L \gg \xi_E(T)$, one should recover the $L$-dependence
of the low-temperature phase of Eq \ref{sgewidthbelow},
so the scaling function $\Phi(x)$ should present the asymptotic
behavior $\Phi(x) \sim x^{1-\zeta_c} $
yielding the temperature dependence of the prefactor
\begin{eqnarray}
\Delta E(L,T) \opsimeq_{L \gg \xi_E(T)} 
\left( \frac{1}{\xi_E(T)} \right)^{1-\zeta_c} L
\label{sgewidthfsslow}
\end{eqnarray}
As shown on Fig \ref{figlogxienerentropiesgb4},
this leads to the same divergence as in Eq \ref{xifreenusg}
\begin{eqnarray}
\xi_E(T) \opsimeq (T_c-T)^{- \nu_{var}} \ \ \ {\rm with } 
\ \ \nu_{var} \simeq 2.95
\label{sgewidthxilow}
\end{eqnarray}

\section{ Summary and conclusions}

\label{conclusion}

In this paper, we have studied the statistical properties
of critical system-size interfaces in a disordered Potts ferromagnet.
For the interface free-energy, our numerical results 
point towards the following
singular behavior for the interface free-energy 
\begin{eqnarray}
\nonumber 
F^{inter}(L,T<T_c)  &&  \simeq 
  f_0(T) L^{d_s} +  \Upsilon(T) L^{\theta} u_F +...
\\  && \oppropto_{T \to T_c^-} 
  \left( \frac{L}{\xi_{av}(T)} \right)^{d_s} 
+   \left( \frac{L}{\xi_{var}(T)} \right)^{\theta} u_F(T,L) +... 
\label{freeconclusion}
\end{eqnarray}
where the average contribution and the random contribution 
involve two correlation lengths $\xi_{av}(T)$ and $\xi_{var}(T)$
that diverge with distinct exponents at criticality (Eq \ref{xifreenupottsq8}).
The 'true' correlation length is expected to be $\xi_{av}(T)$
that appears in the extensive non-random contribution
to the interface free-energy. 
In particular, we have found that the
 interface energy follows the scaling form
\begin{eqnarray}
E^{inter}(L,T<T_c)   \oppropto_{T \to T_c^-} 
\frac{1}{T_c-T} \left[  \left( \frac{L}{\xi_{av}(T)} \right)^{d_s} 
+   \left( \frac{L}{\xi_{av}(T)} \right)^{\frac{d_s}{2}} u_E +...\right]
\label{enerconclusion}
\end{eqnarray}
i.e. both the average contribution and the random contribution 
involve the same correlation length $\xi_{av}(T)$.
This result seems natural within the Fisher-Huse droplet theory
\cite{Fis_Hus,Fis_Hus_DP}
where the interface energy is a sum of random terms that follow
some Central Limit asymptotic behavior.
This picture is confirmed by the Gaussian distribution of the random
variable $u_E$, and by the 'conventional' behavior exactly at criticality
\begin{eqnarray}
E^{inter}(L,T_c)   \sim L^{\frac{1}{\nu_{av}}} u_{E_c}
\label{enertcconclusion}
\end{eqnarray}
However, the presence of another 
length scale $\xi_{var}(T)$ that diverges with a greater exponent
$\nu_{var} > \nu_{av} $ remains to be better understood,
in particular if one compares with the spin-glass transition.
In the spin-glass case, $\xi_{var}(T)$ appearing in the random contribution
of the free-energy is considered as the 'true' correlation length,
since this is the leading term in the free-energy in this case.
But then the interface energy is governed by some critical chaos exponent 
$ E^{inter}(L,T_c) \sim L^{\zeta_c}$ exactly at criticality,
 with $\zeta_c>1/\nu_{var}$, in contrast with
the 'conventional' behavior of Eq. \ref{enertcconclusion}.
So in both cases, even if the physical interpretation 
is different, one needs two different
diverging length scales to describe 
the critical behaviors of the random contributions of the
free-energy and energy or entropy. 
The physical origin seems to be in the 
chaos property of the random variable $u_F(T,L)$ of order $O(1)$
in Eq. \ref{freeconclusion}.
 Within one disordered sample,
the random variable $u_F(T,L)$ strongly depends on the temperature,
and this is why the energy and the entropy of the interface 
presents fluctuations 
that are not directly related to the scalings appearing in the free-energy.
More precisely, the entropy can be obtained as a derivative of the free-energy
with respect to temperature
\begin{eqnarray}
S^{inter}(L,T<T_c)  && = - \frac{ d F^{inter}(L,T)}{ dT} \\
&& = - \frac{ d f_0(T)}{ dT} L^{d_s} - \frac{ d\Upsilon (T)}{ dT} L^{\theta} u_F(T,L)
- \Upsilon(T) L^{\theta} \frac{ \partial u_F(T,L)}{ \partial T}
\label{entropiebilan}
\end{eqnarray}
(and similarly the energy reads $E^{inter} = F^{inter} - T \frac{ d F^{inter}}{ dT}$).
The first term is the extensive term, the second term is only of order $L^{\theta}$,
and thus we conclude that the fluctuation term of order $\sigma(T) L^{d_s/2}$
with $d_s=d-1$ present in the entropy and in the energy ( Eq. \ref{enerdp})
has for origin the derivative of the random variable $u_F(T,L)$.
The identification of these two terms yields
\begin{eqnarray}
\Upsilon(T) L^{\theta} \frac{ \partial u_F(T,L)}{ \partial T} \sim 
\sigma(T) L^{\frac{d_s}{2}} u_E
\label{sigma}
\end{eqnarray}
i.e. the derivative 
\begin{eqnarray}
 T  \frac{ \partial u_F(T,L)}{ \partial T} \sim 
\frac{\sigma(T)}{\Upsilon(T)} L^{\frac{d_s}{2}-\theta} u_E \sim
\frac{\sigma(T)}{\Upsilon(T)} L^{\zeta} u_E
\label{derivative_uf}
\end{eqnarray}
is of order $L^{\zeta}$ where $\zeta=\frac{d_s}{2}-\theta$ 
is the chaos exponent of the zero-temperature fixed point.
Let us now consider the singularity of the prefactor 
$\frac{\sigma(T)}{\Upsilon(T)}$ as $T \to T_c$.
If one defines a chaos length $L_{ch}(T)$ via
\begin{eqnarray}
\frac{\sigma(T)}{\Upsilon(T)} L^{\zeta} \sim  
 \frac{1}{ (T_c-T) } \left( \frac{L}{L_{ch}(T)} \right)^{\zeta} 
\label{deflc}
\end{eqnarray}
one obtains that this chaos length $L_{ch}(T)$
diverges more slowly than the correlation length.
More precisely,
in the random ferromagnetic case, using 
$\Upsilon(T) \sim 1/(\xi_{var}(T))^{\theta} $
and $\sigma(T) \sim 1/((T_c-T) \xi_{av}^{d_s/2}(T))
$, one obtains the singularity
\begin{eqnarray} 
{\rm Random \ Ferromagnets : } \ \ 
L_{ch}(T) = \xi_{av}(T) \times 
 \left( \frac{ \xi_{var}(T)} {\xi_{av}(T)}\right)^{- \frac{\theta}{\zeta}} 
\end{eqnarray}
So here the difference in scaling between the chaos length $L_{ch}(T)$
and the correlation length $\xi_{av}(T)$ comes from the 
difference between $\xi_{var}(T)$ and $\xi_{av}(T)$.
In the spin-glass case, using $\Upsilon(T) \sim 1/(\xi_{var}(T))^{\theta}$
and $\sigma(T) \sim  (\xi_{var}(T))^{\zeta_c-d_s/2}$, 
one obtains the singularity
\begin{eqnarray} 
{\rm Spin \ Glass : } \ \ 
L_{ch}(T) = \xi_{var}(T) \times 
(\xi_{var}(T))^{- \frac{\zeta_c- \frac{1}{\nu_{var}}}{\zeta}}
\end{eqnarray}
Here, the difference in scaling between the chaos length $L_{ch}(T)$
and the correlation length $\xi_{av}(T)$ comes from the 
difference $\zeta_c- \frac{1}{\nu_{var}}>0$.

In conclusion, beyond the differences in interpretation
concerning the nature of the transition in 
disordered ferromagnets   
(where the 'true' correlation length is $\xi_{av}(T)$ 
associated to the non-random term of the free-energy,
and where the interface energy presents conventional scaling at
criticality $E(L,T_c) \sim L^{1/\nu_{av}}$) and in spin-glasses
(where the 'true' correlation length  is $\xi_{var}(T)$ 
associated to the random term of the free-energy,
and where the interface energy presents non-conventional scaling at
criticality $E(L,T_c) \sim L^{\zeta_c}$ with a critical chaos exponent
$\zeta_c>1/\nu_{var}$), the common feature seems to be that in both cases,
the characteristic length scale $L_{ch}(T)$ associated with the chaotic
nature of the low-temperature phase,
diverges more slowly than the correlation length.
Note that for spin-glasses, Nifle and Hilhorst have found
 that the inequality 
$\zeta_c>1/\nu$ is satisfied in a finite range of dimensions $d_-<d<d_+$
above the lower critical dimension $d_-$ ,
whereas for $d>d_+$,
the usual scaling laws in terms of the correlation 
length exponent $\nu$ are valid
\cite{muriel}. In disordered ferromagnets, one may similarly 
wonder whether the different singularities in $\xi_{av}(T)$
and $\xi_{var}(T)$ exist only in a finite range of dimensions $d_-<d<d_+$.
It would be nice to clarify in which conditions
the critical point of a disordered model is described 
by a single diverging length scale or by two diverging length scales. 
This probably requires a more precise understanding 
of the geometrical properties
of the interface in the critical region that should be different for $d<d_+$
and $d>d_+$.

\section*{ Acknowledgements }

It is a pleasure to thank Henk Hilhorst and David Huse
for useful remarks and suggestions.

\appendix

\section{ Reminder on the pure Potts model
 on hierarchical lattices }

\label{pottspure}

In the pure case, the renormalization of Eq. \ref{rgpotts} reduces to the mapping $T$
discussed in \cite{Der_Potts,Der_oscill}
\begin{eqnarray}
x_{n+1}= T(x_n) = \left( \frac{ x_n^2 + (q-1) }
{2 x_n + (q-2)} \right)^b = T(x_n)
\label{rgpottspure}
\end{eqnarray}
The two attractive fixed points $x=1$
(infinite temperature) and $x=\infty$ (zero temperature) are
separated by a repulsive fixed point $x_c$ (critical point).
The critical exponents are obtained as follows 
\cite{Der_Potts,Der_oscill} :
the critical exponent $\nu$ is determined by the linearized mapping
around the critical point
\begin{eqnarray}
\nu = \frac{ \ln 2} {\ln T'(x_c) }
\label{rgpottspurenu}
\end{eqnarray}
and the specific heat exponent reads
\begin{eqnarray}
\alpha= 2- d_{eff} \nu = 2 - \frac{\ln (2b)}{\ln T'(x_c)}
\label{rgpottspurealpha}
\end{eqnarray}
where $d_{eff}= (\ln (2b)/\ln (2)$ represents the dimension
of the hierarchical lattice.
In addition to usual power-laws, there are logarithmic oscillations
coming from the discrete nature of the renormalization \cite{Der_oscill}.
For $b=2$ (effective dimension $d_{eff}=2$), the critical point $x_c(q)$
corresponds to 
\begin{eqnarray}
q=(x_c-1)(\sqrt x_c -1)
\label{xcqb2}
\end{eqnarray}
 and the transition is second order
for any $q$.
The Harris criterion indicates that disorder is relevant
for 
\begin{eqnarray}
q>q_{Harris}(b=2)=4+2 \sqrt{2} \sim  6.828...
\label{harris}
\end{eqnarray}
and this is why we have chosen to use the value $q=8$
for our numerical simulations of the disordered case.


\begin{thebibliography}{99}

\bibitem{Widom}
B. Widom, ''Phase transitions and critical phenomena', Domb and Green Eds, vol. 2, page 79 (NY academic press 1972).

\bibitem{schramm}
O. Schramm,
Israel J. Math. 118, 221 (2000).



\bibitem{sle}
W. Werner, arXiv:math/0303354;
J. Cardy, Ann. Phys. NY 318, 81 (2005);
M. Bauer and D. Bernard, Phys. Rep. 432, 115 (2006). 


\bibitem{gamsa}
A. Gamsa and J. Cardy, J. Stat. Mech. P12009 (2005).

\bibitem{santa}
 M. Picco and R. Santachiara, arXiv:0708.4295.


\bibitem{selke}
W. Selke and W. Resch, Z. Phys. B 47, 335 (1982);
W. Selke and D.A. Huse, Z. Phys. B 50, 113 (1983);
W. Selke, D.A. Huse and D.M. Kroll, J. Phys. A 17, 3019 (1984);
J. Yeomans and B. Derrida,  J. Phys. A 18, 2343 (1985).


\bibitem{Fis_Hus}
D.S. Fisher and D.A. Huse, Phys. Rev. Lett. 56, 1601 (1986);
D.S. Fisher and D.A. Huse, Phys. Rev.  B38, 386 (1988).



\bibitem{heidelberg}
A.J. Bray and M. A. Moore, in Heidelberg colloquium on glassy dynamics, J.L. van Hemmen and I. Morgenstern, Eds (Springer Verlag,
Heidelberg, 1986).

\bibitem{Ban_Bray}
J.R. Banavar and A.J. Bray, Phys. Rev. B 35, 8888 (1987);
T. Aspelmeier, A.J. Bray and M.A. Moore, Phys. Rev. Lett. 89, 197202 (2002).


\bibitem{muriel}
M. Nifle and H.J. Hilhorst, Phys. Rev. Lett. 68 (1992) 2992 ;
M. Ney-Nifle and H.J. Hilhorst, Physica A 193 (1993) 48;
M. Ney-Nifle and H.J. Hilhorst, Physica A 194 (1993) 462;
M. Ney-Nifle, Phys. Rev. B 57, 492 (1998).


\bibitem{thill}
M.J. Thill and H.J. Hilhorst, J. Phys. I France 6, 67 (1996)






\bibitem{Hus_Hen}
D. A. Huse, C. L. Henley, Phys. Rev. Lett. 54, 2708 (1985).

\bibitem{Fis_Hus_DP}
D.S. Fisher and D.A. Huse,  Phys. Rev.  B43, 10728  (1991).


\bibitem{Hus_Hen_Fis}
D. A. Huse, C. L. Henley, and D. S. Fisher, 
Phys. Rev. Lett. 55, 2924 (1985).

\bibitem{Kar}
M. Kardar, Nucl. Phys. B {\bf 290} 582 (1987).

\bibitem{Joh}
K. Johansson, Comm. Math. Phys. 209 (2000) 437.



\bibitem{entropy}
X.H. Wang, S. Havlin and M. Schwartz, Phys. Rev. E 63 (2001) 032601;
X.H. Wang, S. Havlin and M. Schwartz, J. Phys. Chem. B 104 (2000) 3875.

\bibitem{zhang}
Y.C. Zhang, Phys. Rev. Lett. {\bf 59} 2125 (1987);
T.Nattermann, Phys. Rev. Lett. 60 , 2701 (1988).

\bibitem{feigelman}
M.V. Feigelman and V.M. Vinokur, Phys. Rev. Lett. 61 (1988) 1139.

\bibitem{shapir}
Y. Shapir,
Phys. Rev. Lett. 66, 1473 (1991).



\bibitem{diamondcritipolymers}
C. Monthus and T. Garel,  arXiv:0710.0735.

\bibitem{slesg}
C. Amoruso, A.K. Hartmann, M.B. Hastings and M.A. Moore,
Phys. Rev. Lett. 97, 267202 (2006);
D. Bernard, P. Le Doussal and A.A. Middleton, Phys. Rev. B76,
020403(R) (2007). 

\bibitem{fractalSG3d}
S. Risau-Gusman, F. Roma, arXiv:0711.0205.

\bibitem{domany}
S. Wiseman and E. Domany, Phys. Rev. E 52, 3469 (1995);
A. Aharony and A.B. Harris, Phys. Rev. Lett. 77, 3700 (1996);
S. Wiseman and E. Domany, Phys. Rev. Lett. 81, 22 (1998);
S. Wiseman and E. Domany, Phys. Rev. E 58, 2938 (1998).


\bibitem{Kardar_branching}
M. Kardar, A.L. Stella, G. Sartoni and B. Derrida,
Phys. Rev. E 52, R1269 (1995).

\bibitem{Cardy_branching}
J. Cardy, 
Nucl. Phys. B 565, 506 (2000).


\bibitem{realspaceRG}
Th. Niemeijer, J.M.J. van Leeuwen, ''Renormalization theories for Ising
spin systems'' in Domb and Green Eds, ''Phase Transitions and Critical
 Phenomena'' (1976); T.W. Burkhardt and J.M.J. van Leeuwen, 
``Real-space renormalizations'', Topics in current Physics,
 Vol. 30, Spinger, Berlin (1982);
B. Hu, Phys. Rep. 91, 233 (1982).

\bibitem{MKRG}
A.A. Migdal, Sov. Phys. JETP 42, 743 (1976) ; 
L.P. Kadanoff, Ann. Phys. 100, 359 (1976).

\bibitem{berker}
A.N. Berker and S. Ostlund, J. Phys. C 12, 4961 (1979).

\bibitem{hierarchical}
M. Kaufman and R. B. Griffiths, Phys. Rev. B 24, 496 - 498 (1981);
    R. B. Griffiths and M. Kaufman, Phys. Rev. B 26, 5022  (1982).

\bibitem{diluted}
C. Jayaprakash, E. K. Riedel and M. Wortis, 
Phys. Rev. B 18, 2244 (1978)


\bibitem{Kin_Dom}
W. Kinzel and E. Domany, Phys. Rev. B 23, 3421 (1981).

\bibitem{Der_Potts}
B. Derrida and E. Gardner, J. Phys. A 17, 3223 (1984);
B. Derrida in "Critical phenomena, random systems , gauge theories", 
Les Houches 1984, K. Osterwalder
and R. Stora (Eds), North Holland (1986), page 989.

\bibitem{andelman}
D. Andelman and A.N. Berker, Phys. Rev. B 29, 2630 (1984).



\bibitem{young}
A. P. Young and R. B. Stinchcombe,
J. Phys. C 9 (1976) 4419 ; 
B. W. Southern and A. P. Young
J. Phys. C 10 ( 1977) 2179.

\bibitem{mckay}
S.R. McKay, A.N. Berker and S. Kirkpatrick,
Phys. Rev. Lett. 48 (1982) 767;
E. J. Hartford and S.R. McKay, J. Appl. Phys. 70, 6068 (1991).

\bibitem{Gardnersg}
E. Gardner, J. Physique 45, 115 (1984).


\bibitem{bray_moore}
A.J. Bray and M. A. Moore, J. Phys. C 17 (1984) L463;
J.R. Banavar and A.J. Bray, Phys. Rev. B 35, 8888 (1987);
M. A. Moore, H. Bokil, B. Drossel
    Phys. Rev. Lett. 81 (1998) 4252;
S. Boettcher, Eur. Phys. J. B 33, 439 (2003).



\bibitem{diamondtails}
C. Monthus and T. Garel,  arXiv:0710.2198

\bibitem{Der_Gri}
B. Derrida and R.B. Griffiths, Europhys. Lett. 8 , 111 (1989).






\bibitem{Der_oscill}
B. Derrida, C. Itzykson and J.M. Luck,
Comm. Math. Phys. 94, 115 (1984).



\end{thebibliography}
\end{document}